\documentclass[twocolumn]{aastex63}

\usepackage{amsmath}
\usepackage{graphicx}
\usepackage{appendix}
\usepackage{comment}

\newcommand{\isofont}[1]{{\mathrm{#1}}}
\newcommand{\isomass}[1]{{\ensuremath{\isofont{^{#1}}}}}
\newcommand{\isocharge}[1]{{\ensuremath{\isofont{_{#1}}}}}
\newcommand{\isotope}[3]{{\ensuremath{\isocharge{#1}\isomass{#2}\isofont{#3}}}}

\newcommand{\I}[2]{{\isotope{}{#1}{#2}}}

\newcommand{\Ep}[1]{{\ensuremath{10^{#1}}}}
\newcommand{\E}[1]{{\ensuremath{\powersep\Ep{#1}}}}

\newcommand{\powersep}{\times}

\newcommand{\unitstyle}[1]{\ensuremath{\mathrm{#1}}}

\newcommand{\Kelvin}{\unitstyle{K}}
\newcommand{\K}{\Kelvin}  

\newcommand{\Msun}{\ensuremath{\unitstyle{M}_\odot}}
\newcommand{\Lsun}{\ensuremath{\unitstyle{L}_{\odot}}}

\newcommand{\code}[1]{\texttt{#1}}
\newcommand{\mesa}{\code{MESA}}
\newcommand{\MESA}{\mesa}

\newcommand{\GYRE}{\code{GYRE}}

\RequirePackage{bm}

\newcommand{\D}{{\mathrm d}}

\newcommand{\nuclei}[2]{\ensuremath{\mathrm{^{#1}#2}}}

\newcommand{\hydrogen}[1][1]{\nuclei{#1}{H}}
\newcommand{\helium}[1][4]{\nuclei{#1}{He}}

\newcommand{\carbon}[1][12]{\nuclei{#1}{C}}
\newcommand{\nitrogen}[1][14]{\nuclei{#1}{N}}
\newcommand{\oxygen}[1][16]{\nuclei{#1}{O}}

\newcommand{\neon}[1][20]{\nuclei{#1}{Ne}}

\newcommand{\magnesium}[1][24]{\nuclei{#1}{Mg}}

\newcommand{\silicon}[1][28]{\nuclei{#1}{Si}}

\newcommand{\sulfur}[1][32]{\nuclei{#1}{S}}

\newcommand{\iron}[1][56]{\nuclei{#1}{Fe}}

\newcommand{\grada}{\ensuremath{\nabla_{{\rm ad}}}}

\newcommand{\gradT}{\ensuremath{\nabla_T}}
\newcommand{\Teff}{\ensuremath{T_{\rm eff}}}	

\newcommand{\BV}{Brunt-V\"{a}is\"{a}l\"{a}}

\newcommand\rot{\ensuremath{\Omega/\Omega_{{\rm crit}}}}         

\newcommand{\chiT}{\chi_{T}}
\newcommand{\chir}{\chi_{\rho}}

\newlength{\apjcolwidth}
\newlength{\figwidth}
\newlength{\doublewide}
\begin{document}

\title{On The Impact Of $^{22}$Ne On The Pulsation Periods Of Carbon-Oxygen White Dwarfs With Helium Dominated Atmospheres}

\shorttitle{$^{22}$Ne and the Pulsation Periods Of DBV WD}
\shortauthors{Chidester et al.} 

\author[0000-0002-5107-8639]{Morgan T. Chidester}
\affiliation{School of Earth and Space Exploration, Arizona State University, Tempe, AZ 85287, USA}
\affiliation{Joint Institute for Nuclear Astrophysics - Center for the Evolution of the Elements, USA}

\author[0000-0002-0474-159X]{F.X.~Timmes}
\affiliation{School of Earth and Space Exploration, Arizona State University, Tempe, AZ 85287, USA}
\affiliation{Joint Institute for Nuclear Astrophysics - Center for the Evolution of the Elements, USA}

\author[0000-0002-4870-8855]{Josiah Schwab}
\affiliation{Department of Astronomy and Astrophysics, University of California, Santa Cruz, CA 95064, USA}

\author[0000-0002-2522-8605]{Richard H. D. Townsend}
\affiliation{Department of Astronomy, University of Wisconsin-Madison, Madison, WI 53706, USA}

\author[0000-0002-5794-4286]{Ebraheem Farag}
\affiliation{School of Earth and Space Exploration, Arizona State University, Tempe, AZ 85287, USA}
\affiliation{Joint Institute for Nuclear Astrophysics - Center for the Evolution of the Elements, USA}

\author[0000-0002-8107-118X]{Anne Thoul}
\affiliation{Space sciences, Technologies and Astrophysics Research (STAR) Institute, Universit\'e de Li\`ege, All\'ee du 6 Ao$\hat{u}$t 19C, Bat. B5C, 4000 Li\`ege, Belgium}

\author[0000-0002-8925-057X]{C.~E.~Fields}
\affiliation{Department of Physics and Astronomy, Michigan State University, East Lansing, MI 48824, USA}
\affiliation{Joint Institute for Nuclear Astrophysics - Center for the Evolution of the Elements, USA}

\author[0000-0002-4791-6724]{Evan B. Bauer}
\affiliation{Center for Astrophysics $\vert$ Harvard \& Smithsonian, 60 Garden St Cambridge, MA 02138, USA}
\affiliation{Department of Physics, University of California, Santa Barbara, CA 93106, USA}

\author[0000-0002-6748-1748]{Michael H. Montgomery}
\affiliation{Department of Astronomy and McDonald Observatory, University of Texas, Austin, TX 78712, USA}

\correspondingauthor{Morgan T. Chidester}
\email{taylormorgan32@gmail.com}


\begin{abstract}
We explore changes in the adiabatic low-order g-mode pulsation periods of 0.526, 0.560, and 0.729\,\Msun\
carbon-oxygen white dwarf models with helium-dominated envelopes due to the presence, absence, and enhancement of \neon[22]
in the interior. 
The observed g-mode pulsation periods of such white dwarfs are typically
given to 6$-$7 significant figures of precision.
Usually white dwarf models without \neon[22] are fit to the observed periods 
and other properties. 
The root-mean-square residuals to the $\simeq$\,150$-$400~s low-order g-mode periods 
are typically in the range of $\sigma_{\rm rms}$\,$\lesssim$\,0.3~s, 
for a fit precision of $\sigma_{\rm rms}/ P$\,$\lesssim$\,0.3\%.
We find average relative period shifts of $\Delta P/P$\,$\simeq$\,$\pm$\,0.5\% 
for the low-order dipole and quadrupole g-mode pulsations within the observed effective temperature window, 
with the range of $\Delta P/P$ depending on the specific g-mode, 
abundance of \neon[22], effective temperature, and mass of the white dwarf model.
This finding suggests a systematic offset may be
present in the fitting process of specific white dwarfs when \neon[22] is absent.
As part of the fitting processes involves adjusting the
composition profiles of a white dwarf model, our study on the impact of \neon[22] can 
provide new inferences on the derived interior mass fraction profiles.
We encourage routinely including \neon[22] mass fraction profiles, informed by
stellar evolution models, to future generations of white dwarf model fitting processes.

\end{abstract}

\keywords{
Stellar physics (1621); 
Stellar evolution (1599); 
Stellar pulsations (1625); 
White dwarf stars (1799); 
Non-radial pulsations (1117)
         }
         
\section{Introduction} \label{sec:intro}

Photons emitted from stellar surfaces and neutrinos released from
stellar interiors may not directly reveal all that we want to know about the
internal constitution of the stars.  For example, a direct view of the
chemical stratification from the core to the surface is hidden.  
These interior abundance profiles matter: they impact a star's opacity, 
thermodynamics, nuclear energy generation, and pulsation properties.  
The stellar models, in turn, are used to interpret
the integrated light of stellar clusters and galaxies \cite[e.g.,][]{alsing_2020_aa},
decipher the origin of the elements \citep[e.g.,][]{arcones_2017_aa,placco_2020_aa}, 
predict the frequency of merging neutron stars and black holes \citep[][]{giacobbo_2018_aa,farmer_2020_aa,marchant_2020_aa,abbott_2020_aa}, 
and decipher the population(s) of exploding white dwarfs that underlay 
Type Ia supernova cosmology \citep[e.g.,][]{miles_2016_aa,rose_2020_aa}.

Neutrino astronomy, in concert with stellar models, can probe the 
isotopic composition profiles in energy producing regions of the Sun \citep{borexino-collaboration_2018_aa,borexino-collaboration_2020_aa}
and nearby ($d$\,$\lesssim$\,1\,kpc) presupernova massive stars up to tens of hours before core-collapse
\citep[e.g.,][]{patton_2017_ab,simpson_2019_aa,mukhopadhyay_2020_aa}.
Stellar seismology, also in concert with stellar models, can probe the
elemental composition profiles in pulsating stars
from the upper main-sequence \citep[e.g.,][]{simon-diaz_2018_aa,pedersen_2019_aa,balona_2020_aa}
through the red-giant branch \citep[e.g.,][]{hekker_2017_aa,hon_2018_aa}
to white dwarfs \citep[WDs, e.g.,][]{hermes_2017_ab,giammichele_2018_aa,corsico_2019_aa,bell_2019,bischoff-kim_agnes_2019,althaus_2020_aa}.

Most of a main-sequence star's initial metallicity $Z$ comes from the carbon-nitrogen-oxygen (CNO) 
and \iron[56] nuclei inherited from its ambient interstellar medium. 
All of the CNO piles up at \nitrogen[14] when H-burning on the main-sequence is completed
because the \nitrogen[14]($p$,$\gamma$)\oxygen[15] reaction rate is the slowest
step in the H-burning CNO cycle.  During the ensuing He-burning
phase, all of the $^{14}$N is converted to $^{22}$Ne by the reaction sequence
\nitrogen[14]($\alpha$,$\gamma$)$^{18}$F(,$e^{+}\nu_e$)$^{18}$O($\alpha$,$\gamma$)\neon[22].
The abundance of \neon[22] when He-burning is completed is thus proportional to 
the initial CNO abundance of the progenitor main-sequence star. 
The weak reaction in this sequence powers the neutrino
luminosity during He-burning \citep[e.g.,][]{serenelli_2005_aa,farag_2020_aa}
and marks the first time in a star's life that the core becomes neutron rich.
For zero-age main sequence (ZAMS) masses between
$\simeq$\,0.5\,\Msun \ \citep{demarque_1971_aa, prada-moroni_2009_aa, gautschy_2012_aa}
and $\simeq$\,7\,\Msun\ 
\citep{becker_1979_aa, becker_1980_aa, garcia-berro_1997_aa}, depending 
on the treatment of convective boundary mixing \citep{weidemann_2000_aa, denissenkov_2013_aa,
jones_2013_aa, farmer_2015_aa, lecoanet_2016_aa, constantino_2015_aa, constantino_2016_aa, constantino_2017_aa},
the \nitrogen[14]($\alpha$,$\gamma$)$^{18}$F(,$e^{+}\nu_e$)$^{18}$O($\alpha$,$\gamma$)\neon[22] reaction sequence determines 
the \neon[22] content of a resulting carbon-oxygen white dwarf (CO WD). We follow the convention that 
\neon[22] is the ``metallicity'' of the CO WD.

\citet{camisassa_2016_aa} analyze the impact of \neon[22] on the
sedimentation and pulsation properties of H-dominated atmosphere WDs 
(i.e., the DAV class of WD) with masses of 0.528, 0.576, 0.657, 
and 0.833\,$\Msun$. These WD models result from $Z$\,=\,0.02
non-rotating evolutionary models that start from the ZAMS and
are evolved through the core-hydrogen and core-helium burning, 
thermally pulsing asymptotic giant branch (AGB), and post-AGB phases.  
At low luminosities, $\log(L/\Lsun)$\,$\lesssim$\,$-4.5$, they find that 
\neon[22] sedimentation delays the cooling of WDs by 0.7 to 1.2 Gyr,
depending on the WD mass.
They also find that \neon[22] sedimentation induces differences in the periods that are 
larger than the present observational uncertainties.

\citet{giammichele_2018_aa} analyze in their supplemental material the effect 
of \neon[22] on the pulsation periods of a 0.570 $\Msun$ template-based model
for the DB WD KIC~08626021. 
They considered a model consisting of pure oxygen core surrounded by a pure helium envelope
with the same mass and effective temperature equal to those inferred for KIC~08626021. 
Next, they considered a model that replaces the pure oxygen core with an oxygen-dominated core plus a trace
amount of \neon[22]. They find that the model with \neon[22] has, on average, shorter pulsation periods.

This article is novel in three ways.
One, we explore the impact of \neon[22] on the adiabatic low-order g-mode pulsation periods of 
CO WD models with a He-dominated atmosphere (i.e., the DBV class of WD) as the models cool
through the range of observed DBV effective temperatures.
Two, we derive an approximation formula for the \BV\ frequency in WDs that allows new physical insights
into why the low-order g-mode pulsation periods change due to the presence, and absence, of \neon[22].
Three, we analyze how the \neon[22] induced changes in the pulsation periods depend on the mass and temporal resolutions of the WD model.
Our explorations can help inform inferences about the interior mass fraction profiles 
derived from fitting the observed periods 
of specific DBV WDs 
\citep[e.g.,][]{metcalfe_2002_aa,fontaine_2002_aa,metcalfe_2003_aa,metcalfe_2003_ab,hermes_2017_ab,
giammichele_2017_aa,giammichele_2018_aa,charpinet_2019_aa,de-geronimo_2019_aa,bischoff-kim_agnes_2019}.

In Section~\ref{s.models} we summarize the input physics, and discuss in detail the chemical stratification, cooling properties,
and g-mode pulsation periods of one DBV WD model.
In Section~\ref{s.pchanges} we present our results on changes to the low-order g-mode pulsation periods
due to the presence, or absence, of \neon[22] from this model. 
In Section~\ref{s.others} we study changes in the 
g-mode pulsation periods due to \neon[22] from a less massive and a more massive WD model. 
In Section~\ref{s.summary} we summarize and discuss our results.
In Appendix~\ref{sec:convergence} we study the robustness of our results
with respect to mass and temporal resolution, and in Appendix~\ref{sec:details}
we discuss in more depth some of the input physics.

\begin{figure*}[!htb]
\centering
\includegraphics[width=1.8\apjcolwidth]{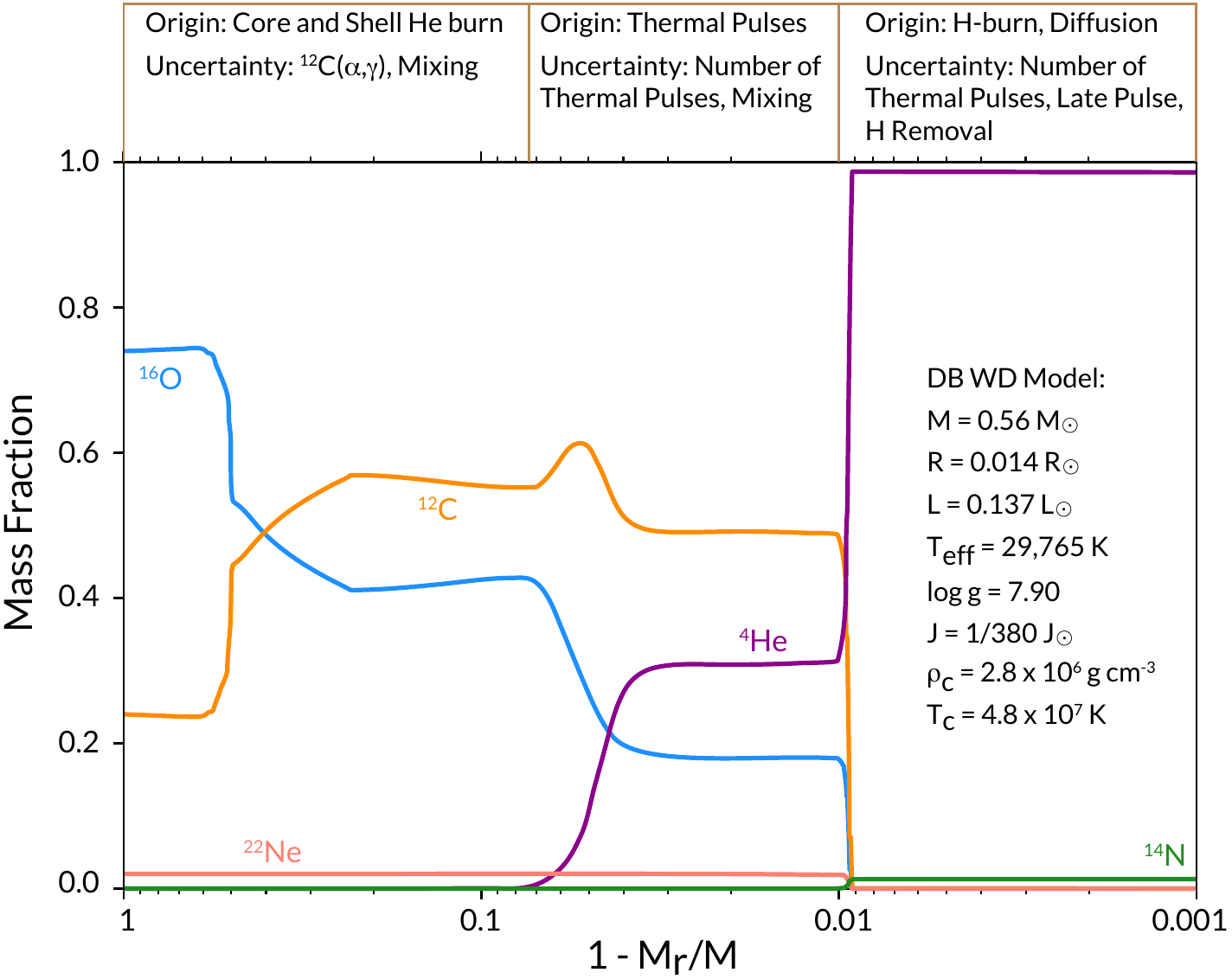}
\caption{
Mass fraction profiles of the 0.56\,\Msun DB WD resulting from the evolution of the 2.1\,\Msun, $Z$=0.02, ZAMS
model. 
}
\label{fig:illus}
\end{figure*}

\section{A Baseline WD Model} \label{s.models}

\subsection{Input Physics}\label{s.input_physics}

We use \MESA\ version r12115 
\citep{paxton_2011_aa,paxton_2013_aa,paxton_2015_aa,paxton_2018_aa,paxton_2019_aa}
to evolve a $2.1\,\Msun$, $Z$\,=\,0.02 metallicity model from the ZAMS 
through core H-burning and core He-burning.
After winds during the thermal pulses on the AGB have reduced the H-rich
envelope mass to $0.01\,\Msun$, the remaining hydrogen is stripped
from the surface to form a young, 0.56~\Msun DB WD.  
This model is tuned to match the observed and inferred properties of KIC~08626021
\citep{bischoff-kim_ostensen_2014,giammichele_2018_aa,timmes_2018_aa,charpinet_2019_aa,de-geronimo_2019_aa}.
Additional details of the input physics are given in Appendix~\ref{sec:details}, and 
the \MESA\ r12115 files to reproduce our work are available
at \dataset[https://doi.org/10.5281/zenodo.4338180]{https://doi.org/10.5281/zenodo.4338180}


\subsection{Mass Fraction Profiles}\label{s.massfractions}

Figure \ref{fig:illus} shows the mass fraction $X(^AZ)$ profiles of the resulting 0.56\,\Msun DB WD model,
where $A$ is the number of nucleons and $Z$ is the number of protons.
Brown boxes divide the mass fraction profiles into three regions
according to their origins and uncertainties.  The $X(\carbon[12])$ 
and $X(\oxygen[16])$ profiles in the innermost $\simeq$\,90\% by mass region are determined
during core and shell He-burning. The main uncertainties in this
region are the \carbon[12]($\alpha$,$\gamma$)\oxygen[16] reaction rate
\citep[e.g.,][]{deboer_2017_aa}, and the treatment of convective mixing
boundaries during core H-and He-burning
\citep[e.g.,][]{constantino_2015_aa,constantino_2016_aa,constantino_2017_aa}.

The CO and $X(\helium[4])$ profiles between $\simeq$\,1\% and $\simeq$\,10\% of
the exterior WD mass originate from shell He-burning during the thermal pulse
phase of evolution on the AGB.  Most of the total He mass 
is held in this region.  
The primary uncertainties in this region
are the number of thermal pulses and convective boundary layer mixing. The number of
thermal pulses a model undergoes is sensitive to the mass
resolution, time resolution, mass loss rate, and the treatment of
convective boundaries \citep{iben_1983_aa,herwig_2005_aa,karakas_2014_aa}.  
The sharp change in all the mass
fractions at $\simeq$\,1\% of the exterior WD mass marks the extent reached by
the convective boundary during the last thermal pulse. 

CO profiles in this region may also be subject to other mixing
processes. For example, the magnitude of the $X(\carbon[12])$ ``bump'' 
is subject to the strength and duration of the
thermohaline instability, which occurs when material is stable to
convection according to the Ledoux criterion, but has an inverted 
molecular weight gradient \citep{baines_1969_aa,brown_2013_ab,garaud_2018_aa,bauer_2018_aa}.

The $X(\helium[4])$ profile of the outer $\simeq$\,0.1\% to 1\% of the WD
mass is determined by shell H-burning. All of the initial CNO mass
fractions have been converted to \nitrogen[14]. The main uncertainties
in this region are the number of thermal pulses during the AGB phase
of evolution, late or very late thermal pulses
\citep{blocker_1995_aa,blocker_1995_ab,blocker_2001_aa}, and
mechanisms to remove the residual high entropy, H-rich layer to produce
a DB WD from single and binary evolution
\citep[e.g.,][]{dantona_1990_aa,althaus_1997_aa,parsons_2016_aa}.

The $X(\neon[22])$ profile is essentially flat and spans the inner $\simeq$\,99\% by mass.
As discussed in Section~\ref{sec:intro}, $X(\neon[22])$ is created from \nitrogen[14] during He-burning.

\begin{figure*}[!htb]
\centering
\includegraphics[trim={2.0cm 3.5cm 1.8cm 2.5
cm},clip,width=1.8\apjcolwidth]{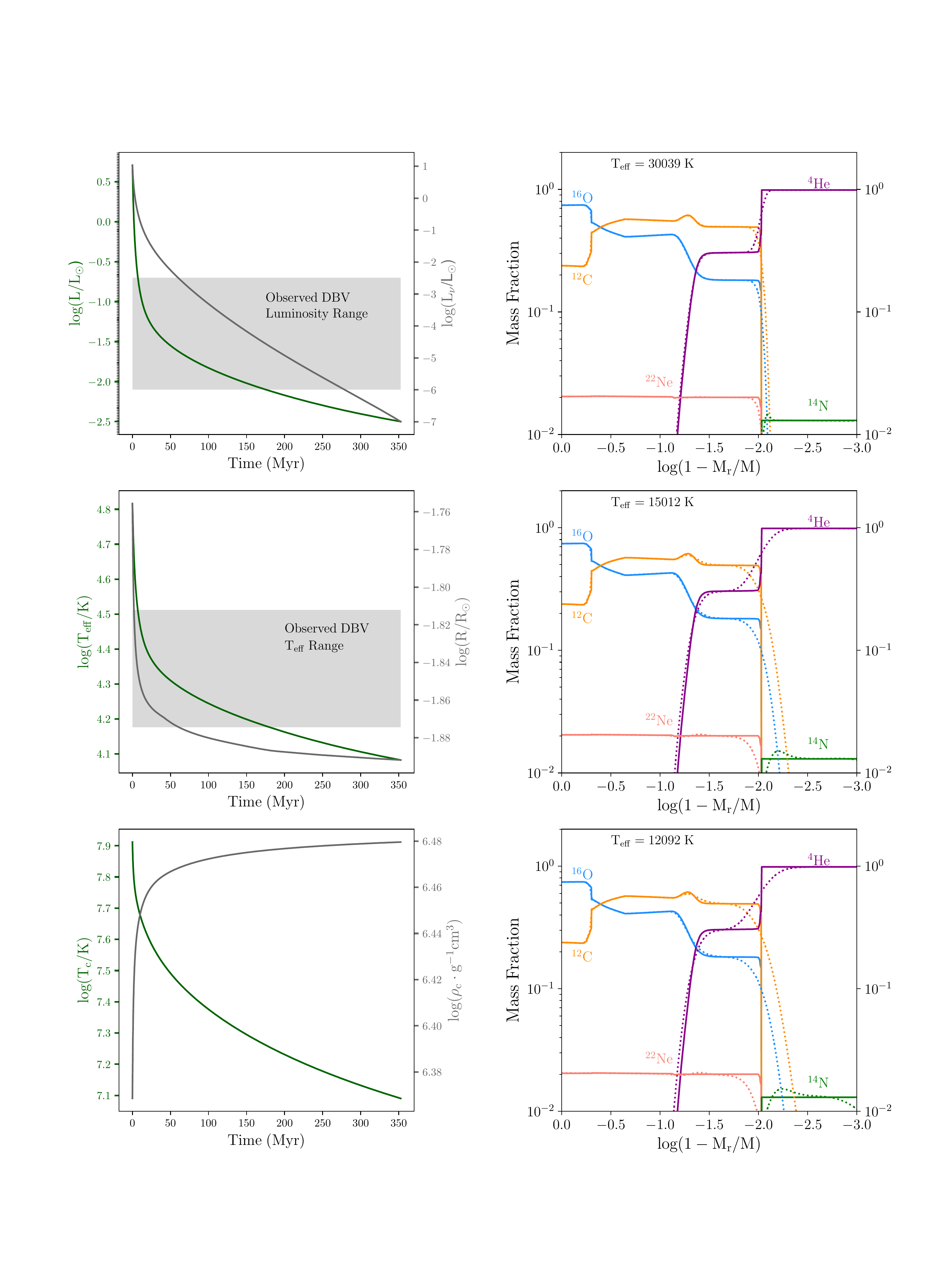}
\caption{
Evolution of baseline model's
photon luminosity $L$ and neutrino luminosity $L_{\nu}$ (left top),
effective temperature $\Teff$ and radius $R$ (left middle),
central temperature $T_c$ and central density $\rho_c$ (left bottom).
Time begins a few thermal timescales after the ab initio WD is released.
Gray bands show the luminosity and $\Teff$ 
range of currently observed DBV WD \citep[Montreal White Dwarf Database,][]{dufour_2017_aa}.
Mass fraction profiles are shown at \Teff\,=\,30,039~K (right top), 15,012~K (right middle),
and 12,092~K (right bottom) and at the end of the evolution.
Initial mass fraction profiles are shown as solid curves 
and the diffusing mass fraction profiles are shown as dotted curves.
}
\label{fig:base1st_last}
\end{figure*}

\begin{figure}[!htb]
\centering
\includegraphics[trim={0.3cm 0.1cm 1.3cm 1.2cm},clip,width=1.0\apjcolwidth]{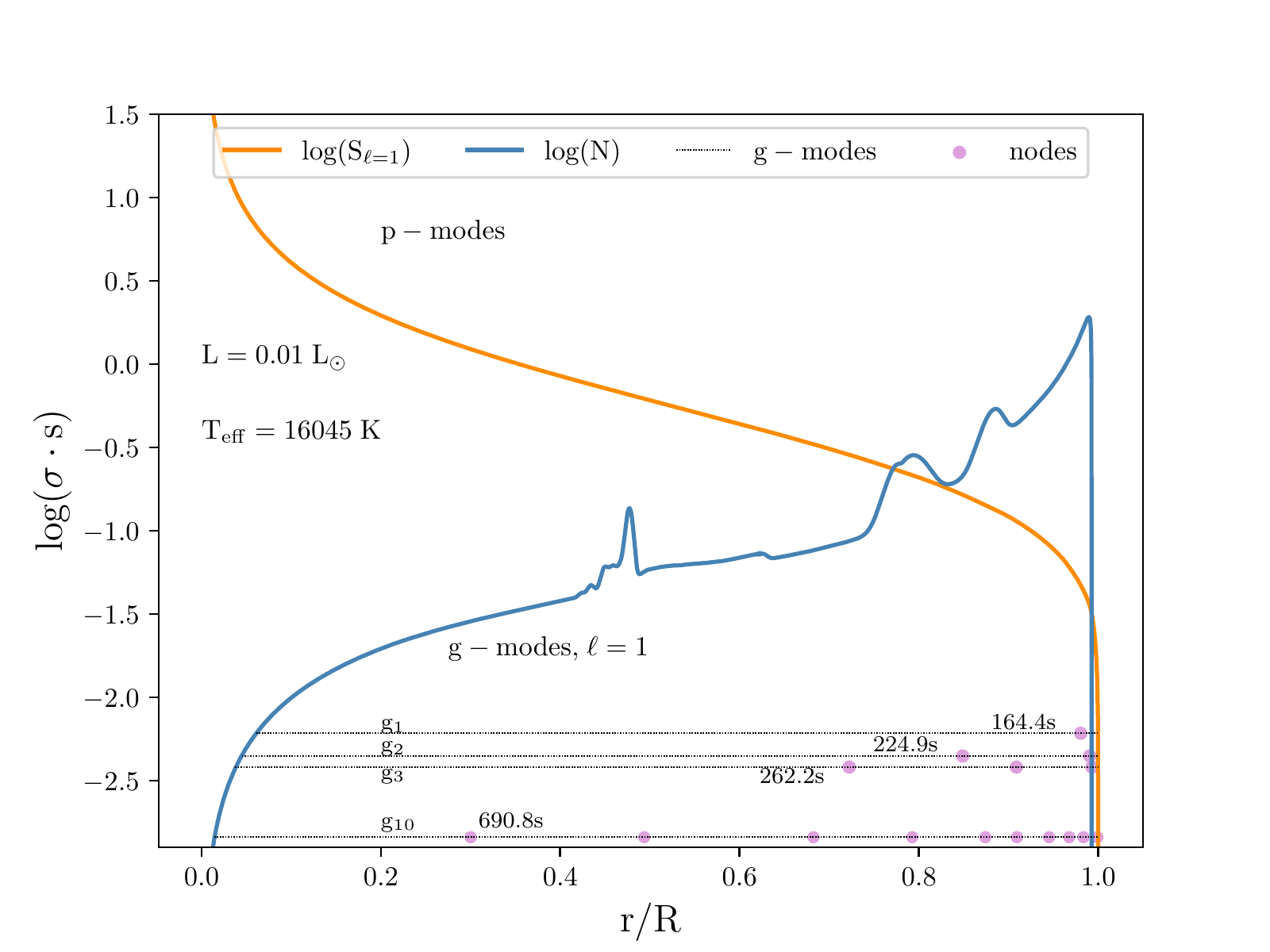}
\includegraphics[trim={0.3cm 0.1cm 1.3cm 1.2cm},clip,width=1.0\apjcolwidth]{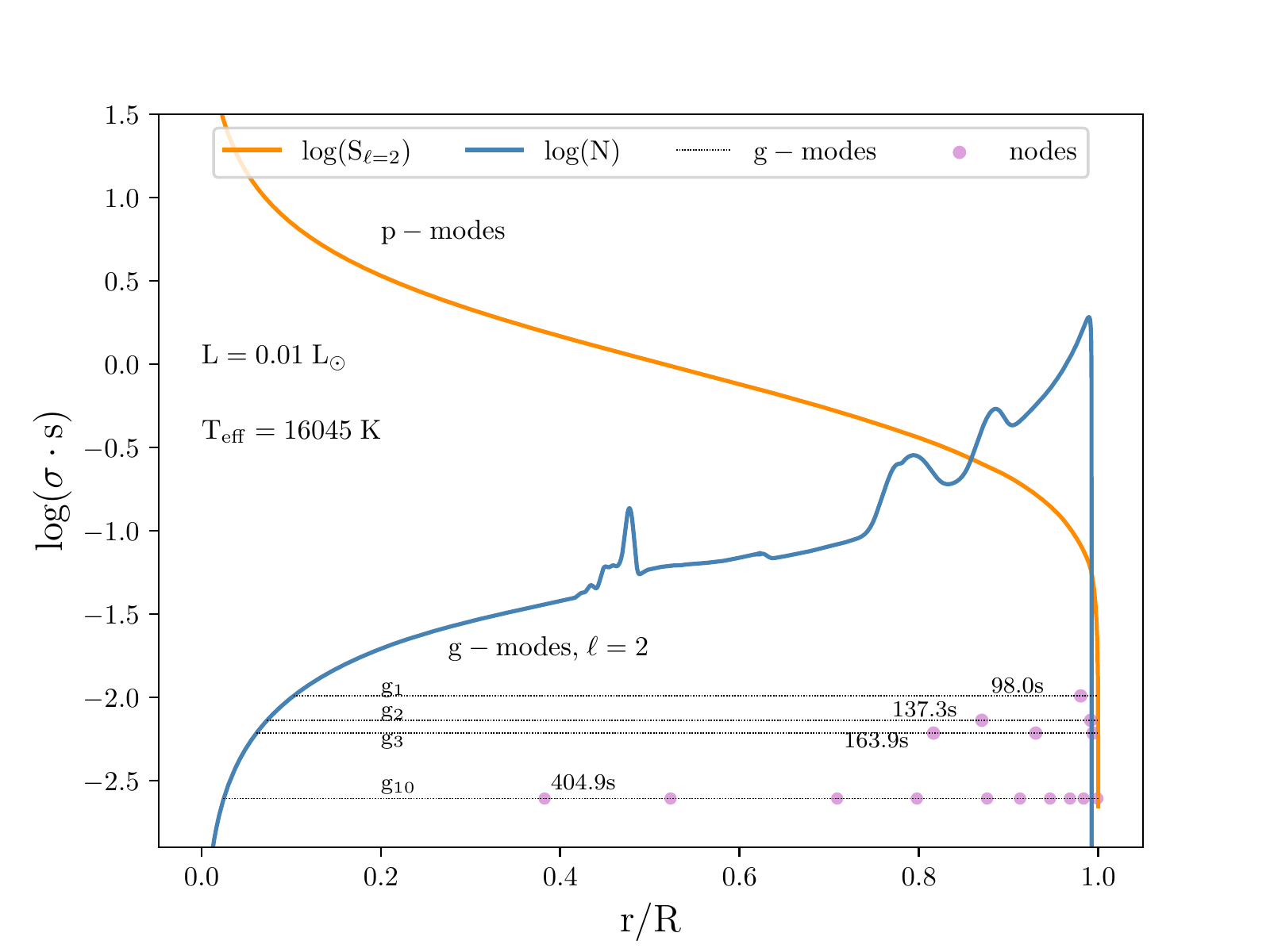}
\caption{
Propagation diagram for the dipole $\ell$\,=\,1 (top) and  quadrupole $\ell$\,=\,2 (bottom)
g-modes at $L$\,=\,0.01\,\Lsun \ for the baseline WD model. The Lamb frequency ($S_{\ell}$, orange), \BV\ frequency ($N$, blue),
radial order $n$\,=\,1,2,3,10 eigenfrequencies (dotted black lines),
nodes in the radial eigenfunction (filled circles), and g-mode 
period of each radial order are labeled.
        }
\label{fig:prop_gmodes}
\end{figure}

\vspace{0.3in}
\subsection{Constructing Ab Initio White Dwarf Models}\label{s.wdbuilder}

Starting from a set of pre-main sequence (pre-MS) initial conditions,
accurate predictions for the properties of the resulting WD model, 
especially the mass fraction profiles, do not exist due to the nonlinear 
system of equations being approximated. 
In addition, evolving a stellar model from the pre-MS to a WD 
can be resource intensive. It can thus be useful for
systematic studies to build ab initio WD models \citep[e.g., WDEC,][]{bischoff-kim_2018_aa}.
By ab initio we mean calculations that begin with a WD model, as opposed to a WD model 
that is the result of a stellar evolution calculation from the pre-MS.
A potential disadvantage (or advantage) of ab initio WD models is the imposed initial mass fraction profiles
may not be attainable by a stellar model evolved from the pre-MS.
Throughout the remainder of this article we use a new capability, \code{wd\_builder},
to construct ab initio WD models in \MESA\ of a given mass and chemical stratification.

The initial structure of an ab initio WD model is 
approximated as an isothermal core and a radiative envelope in hydrostatic equilibrium.
Here we specify an initial WD mass of 0.56~\Msun, the same WD mass as 
produced by the stellar evolution calculation.
The imposed $X(\helium)$, $X(\carbon)$, $X(\nitrogen)$, 
$X(\oxygen)$, and $X(\neon[22])$ profiles are taken from the
stellar evolution derived mass fraction profiles of Figure~\ref{fig:illus}
and normalized to sum to unity in each cell.
Henceforth we refer to this ab initio WD model as the ``baseline model''.

For ab initio WD models we use He-dominated, $\log$(H/He)\,=\,$-$5.0, model atmosphere tables
spanning 5,000\,K\,$\le$\,$\Teff$\,$\le$\,40,000\,K that were provided
by Odette Toloza (2019, private communication) using the Koester WD
atmosphere software instrument \citep{koester_2010_aa}.  These
tabulated atmospheres for DB WDs are publicly available as a standard
atmosphere option as of \MESA\ r12115. In addition,
we use five element classes for the diffusion classes -- \helium[4], 
\carbon[12], \nitrogen[14] \oxygen[16], and \neon[22].
Otherwise, all of the physics implementations and modeling choices are
as described in Section~\ref{s.input_physics}.

The initial baseline model is then evolved with \MESA. 
As the model is not in thermal equilibrium, there is an initial transient phase lasting a
few thermal timescales that is disregarded. The thermal timescale is 
$\tau_{\rm th}$\,$\simeq$\,$E_{\rm th} / L_{\rm tot}$\,$\simeq$\,0.67 Myr,
where  $E_{\rm th}$ is the thermal energy of the WD 
and $L_{\rm tot}$ is the photon plus neutrino luminosity.  Specifically, we set 
the zero point to be 1.5 thermal timescales ($\simeq$\,~1 Myr) after the transient reaches its peak luminosity.
The evolution terminates when $L_{\rm tot}$ falls below $\log(L/\Lsun)$\,=\,$-$2.5.

Figure \ref{fig:base1st_last} shows the cooling properties of the baseline model.
Plasmon neutrino emission dominates the energy loss budget at 
$\Teff \gtrsim 25,000\,{\rm K}$ \citep[e.g.,][]{vila_1966_aa, kutter_1969_aa,winget_2004_aa, bischoff-kim_2018_aa}.  
Photons leaving the
WD surface begin to dominate the cooling as the electrons transition
to a strongly degenerate plasma \citep{van-horn_1971_aa}.
The luminosity becomes proportional to the enclosed mass, $L_r$\,$\propto$\,$\,M_r$,
in this model only when $\Teff \lesssim 20,000\,{\rm K}$ \citep{timmes_2018_aa}.
Energy transport in the interior is dominated by conduction, driven
primarily by electron-ion scattering. Energy transport in the outer
layers is dominated by radiation or convection associated with the
partial ionization of He at $\Teff \simeq 30,000 \, {\rm K}$.

Figure \ref{fig:base1st_last} also shows the 
diffusion of the initial mass fractions
as the baseline WD model cools to
\Teff\,=\,30,000~K, 15,000~K and 12,138~K (corresponding to the termination at $\log(L/\Lsun)$\,=\,$-$2.5).
Element diffusion of \neon[22] is modest for the baseline 0.56\,\Msun\ DB WD model.
Depletion of the \neon[22] mass fraction at $\log(1 - M_r/M$)\,$\simeq$\,$-1.9$ 
has occurred by the time the model has cooled to \Teff\,$\simeq$\,30,000~K.
As the model cools further, the surface regions in the tail of the He-dominated layer
further deplete and a small \neon[22] bump forms and propagates inwards toward the center.
The timescale for
\neon[22] to travel from near the surface to the center of this WD model is
$\tau_{\rm D} \simeq 2 \bar{Z} \Gamma^{1/3} \rho_6^{-1/2} \ {\rm Gyr} \simeq 30 \ {\rm Gyr}$
\citep{isern_1991_aa, bravo_1992_aa, bildsten_2001_aa, deloye_2002_aa, camisassa_2016_aa},
where $\bar{Z}$ is the mean charge of the material, $\Gamma$ is the electrostatic to
thermal energy ratio, and $\rho_6$ is the baryon mass density in units
of 10$^6$~g\,cm$^{-3}$. Thus, the $X(\neon[22])$ profile does not significantly change
as the 0.56~\Msun baseline model evolves to $\log(L/\Lsun)$\,=\,$-$2.5 in $\simeq$\,350 Myr.
More massive WDs show larger amounts of \neon[22] sedimentation over the same time period \citep{camisassa_2016_aa}.
WD cooling data suggests a significant enhancement due to to \neon[22] diffusion
\citep{cheng_2019_aa,bauer_2020_aa}, but does not effect the baseline model 
until it cools to effective temperatures lower than considered here
($\Teff$\,$\lesssim$\,10,000~K).

\subsection{Pulsation Periods of the Baseline Model}\label{s.abperiods}

Having established the structural and composition profiles of a
cooling baseline WD model, we now consider the g-mode
pulsation periods.  Some of the material is classic
\citep[e.g.,][]{unno_1989_aa, fontaine_2008_aa}, but we also derive
and verify the accuracy of an approximation formula for the
\BV\ frequency in WDs that allows physical insights into why the
low-order g-mode pulsation periods change due to variations in the 
mass fraction of \neon[22].  This material is essential for
establishing that the baseline model, before introducing any
modifications to the chemical profiles, produces pulsation periods that are commensurate with
the observations of DBV WDs.

Figure~\ref{fig:prop_gmodes} shows the propagation diagram
\citep[e.g.,][]{unno_1989_aa} for the baseline WD model after it has
cooled to \Teff\,=\,16,045~K and dimmed to $L$\,=\,0.01 \Lsun, within the DBV WD observation window.  
Adiabatic pulsation frequencies are calculated
using release 5.2 of the \GYRE\ software instrument
\citep{townsend_2013_aa,townsend_2018_aa}.  For a fixed radial overtone number, 
the $\ell$\,=\,1 periods
are $\sim$\,$\sqrt{3}$ longer than the
$\ell$\,=\,2 periods, due to the local dispersion relation for low-frequency g-modes
$\sigma_g$ scaling as

\begin{equation}
\sigma_g^2 \simeq \ell(\ell+1)N^2 /(k_r^2 r^2) \ ,
\label{eq:gmodefrequency}
\end{equation}
\noindent
where $k_r$ is the radial wave number. 
The \BV \ frequency $N$ is

\begin{equation}
N^2 = \frac{g^2 \rho}{P} \frac{\chiT}{\chir} (\grada - \gradT + B) \ ,
\label{eq:bv_full}
\end{equation}
where
$g$ is the gravitational acceleration,
$\rho$ is the mass density,
$P$ is the pressure,
$T$ is the temperature,
$\chiT$ is the temperature exponent $\partial({\rm ln}P)/\partial({\rm ln}\rho)|_{T,\mu_I}$,
$\chir$ is the density exponent $\partial({\rm ln}P)/\partial({\rm ln}T)|_{\rho,\mu_I}$,
$\grada$ is the adiabatic temperature gradient,
$\gradT$ is the actual temperature gradient, and
$B$ accounts for composition gradients \citep[e.g.,][]{hansen_1994sipp.book.....H,fontaine_2008_aa}. 
Bumps in the $N$ profile of Figure~\ref{fig:prop_gmodes} correspond to transitions in the 
$X(\oxygen)$, $X(\carbon)$, and $X(\helium)$ profiles.
The implementation of Equation~\ref{eq:bv_full} in \MESA\ is described in 
Section 3 of \citet{paxton_2013_aa}.

An approximation for $N^2$ in the interiors of WDs that yields physical insights 
begins by assuming $\grada$ is much larger than $\gradT$ and $B$. Then 

\begin{equation}
N^2 = \frac{g^2 \rho}{P} \frac{\chiT}{\chir} \grada
\ .
\label{eq:bv_mid1}
\end{equation}
In the interior of a WD the ions are ideal and dominate 
the temperature derivatives of an electron degenerate plasma.
Substituting the pressure scale height $H$\,=\,$P /(\rho g)$ 
and equation 3.110 of \citet{hansen_1994sipp.book.....H}

\begin{equation}
\chiT = \frac{\rho}{P} \frac{k_B T}{\mu_I m_p}
\label{eq:bv_chit}
\end{equation}

\noindent
into Equation~\ref{eq:bv_mid1} gives

\begin{equation}
N^2 = \frac{1}{H^2 \chir} \frac{k_B T}{\mu_I m_p} \grada
\ ,
\label{eq:bv_mid2}
\end{equation}
where 
$k_B$ is the Boltzmann constant,
$\mu_I$\,=\,1/($\sum_i X_i/A_i)$ is the ion mean molecular weight, and
$m_p$ is the mass of the proton.
Equation 3.90 of \citet{hansen_1994sipp.book.....H}
shows
$\grada$\,=\,$(\Gamma_3 - 1)/ \Gamma_1$, where 
$\Gamma_1$ is the first adiabatic index and
$\Gamma_3$\,$\rightarrow$\,$k_B / (\mu_I m_p c_v)$ is the third adiabatic index,
where in the gas phase the ideal specific heat capacity is $c_v$\,=\,$3 k_B/(2 \mu_I m_p)$. 
The sentence beneath 
equation 3.112  of \citet{hansen_1994sipp.book.....H}
thus notes that $\Gamma_3 - 1$\,=\,2/3
for the ions in the gas phase ($\Gamma_3 - 1$\,=\,1/3 in the liquid phase).
Combining these expressions, yields the approximation

\begin{equation}
N^2 = \frac{2}{3 \Gamma_1 \chir H^2} \frac{k_B T}{\mu_I m_p}
\ .
\label{eq:bv_simple}
\end{equation}

\begin{figure}[!htb]
\centering
\includegraphics[trim={0.3cm 0.1cm 1.3cm 1.2cm},clip,width=1.0\apjcolwidth]{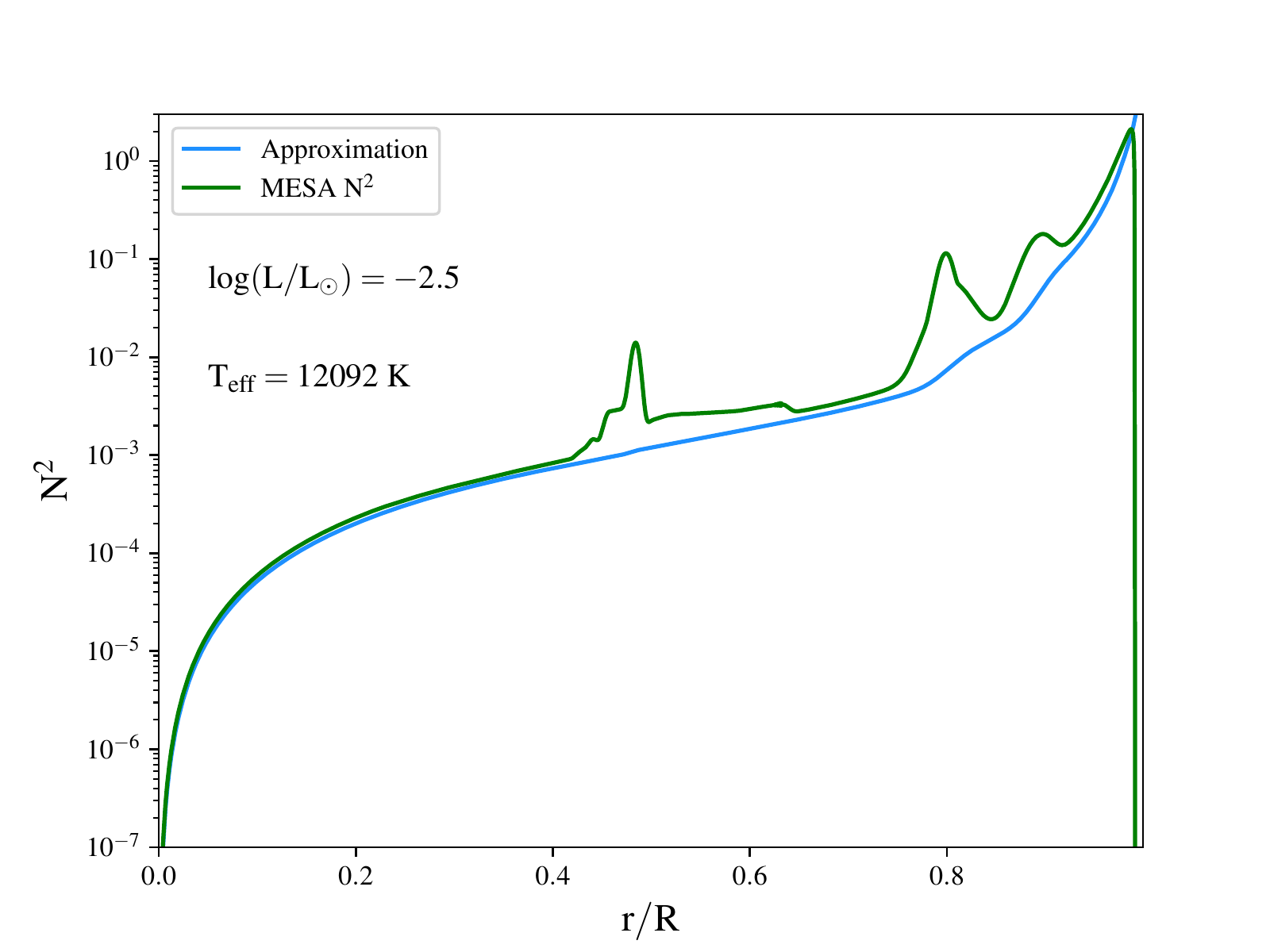}
\caption{Comparison of the approximation for $N^2$ (blue curve) in Equation~(\ref{eq:bv_simple})
and the full calculation of $N^2$ from \MESA\ (green curve).
 }
\label{fig:N2equation1}
\end{figure}

\begin{figure}[!htb]
\centering
\includegraphics[trim={0.3cm 0.1cm 1.3cm 1.2cm},clip,width=1.0\apjcolwidth]{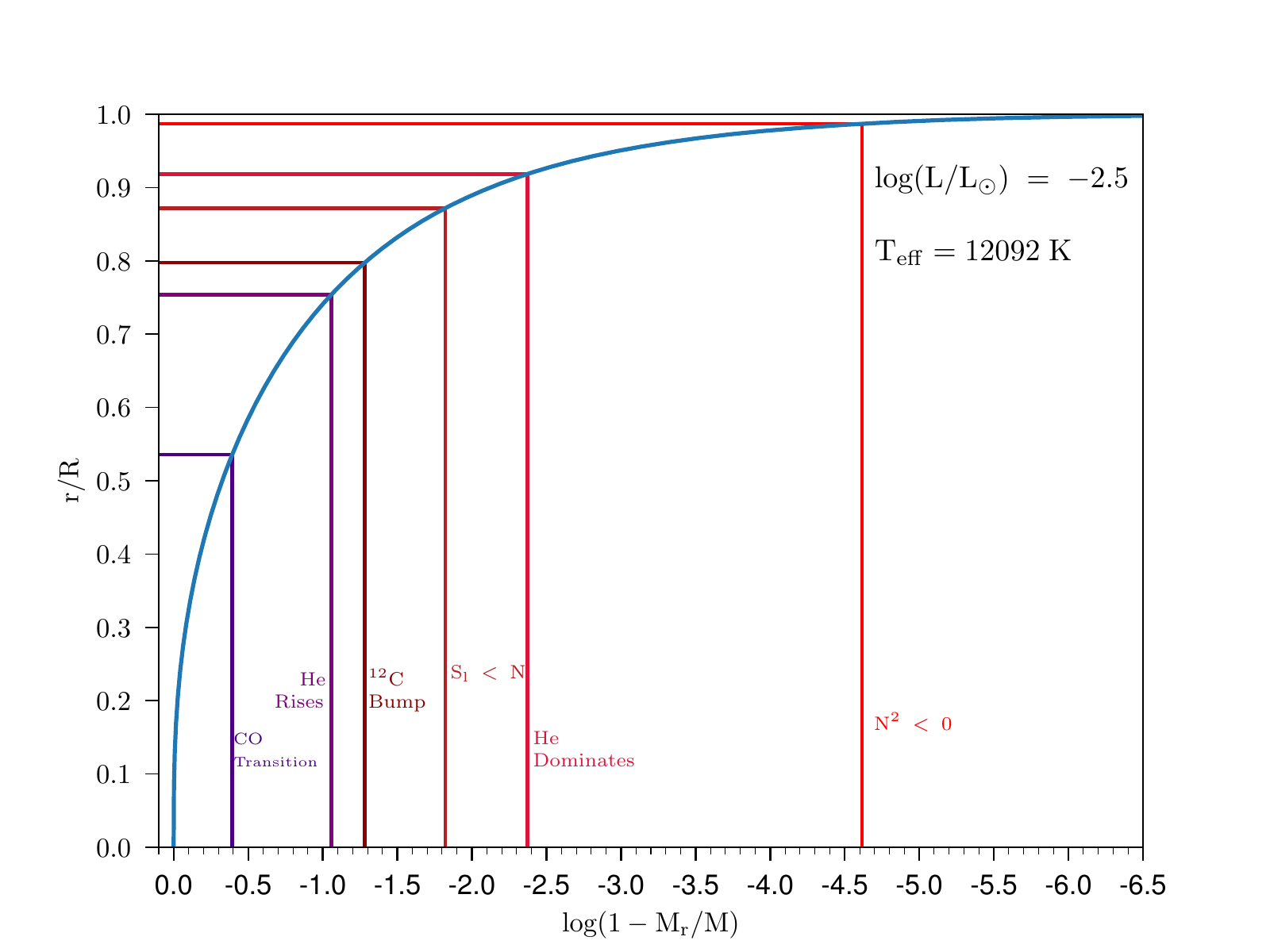}
\caption{
Mass-radius relation of the baseline DB WD model at $\log(L/\Lsun)$\,=\,$-$2.5 with key features located:
the transition from $X(\oxygen[16])$ to $X(\carbon[12])$ dominated,
the rise of $X(\helium)$,
the $X(\carbon)$ bump,
where $S_{\ell}$\,$<$\,$N$ occurs,
the transition to $X(\helium)$ dominated, and
where $N^2$\,$<$\,0.
}
\label{fig:mr}
\end{figure}

\begin{figure}[!htb]
\centering
\includegraphics[trim={0.3cm 0.1cm 1.3cm 1.2cm},clip,width=1.0\apjcolwidth]{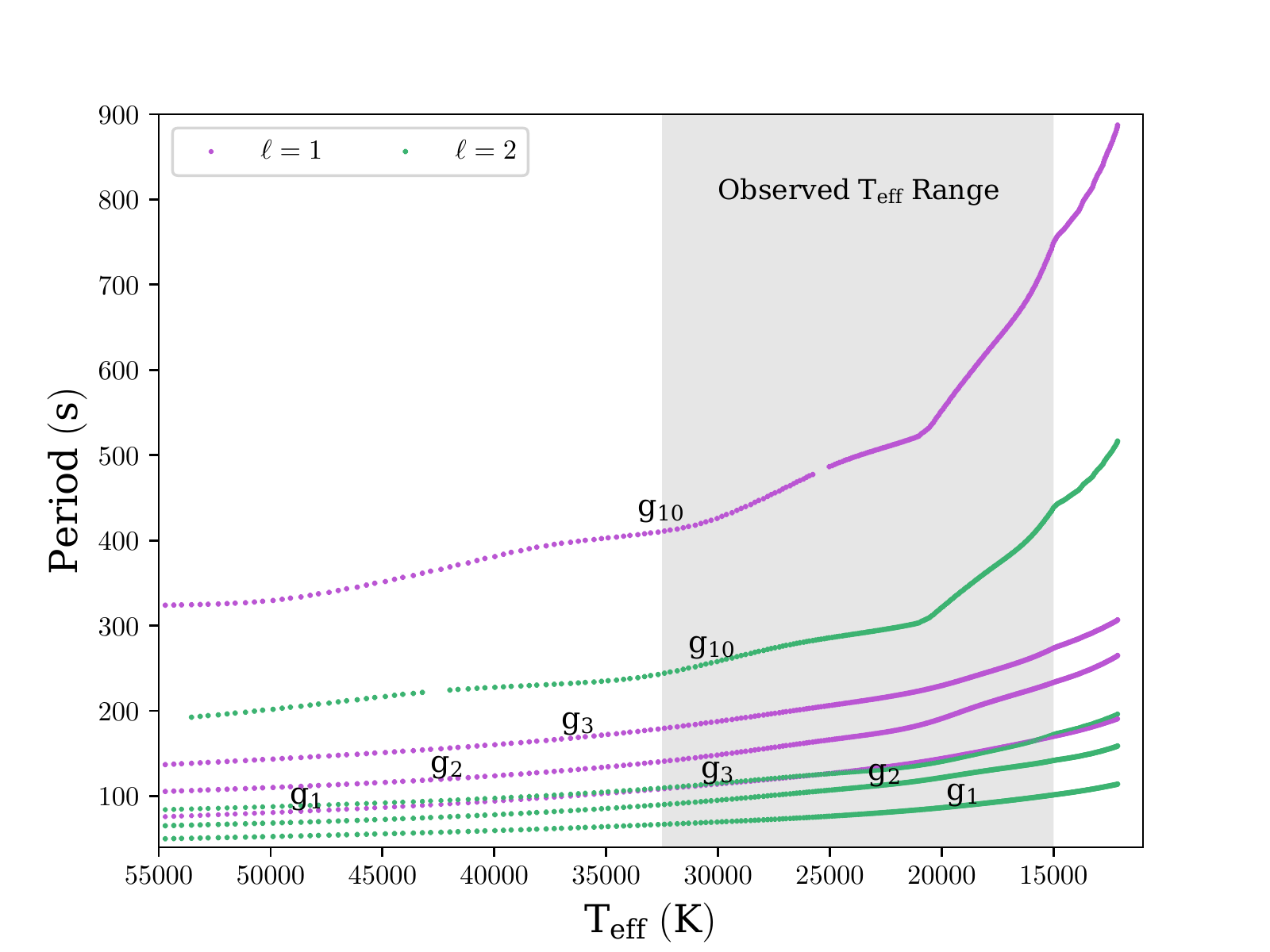}
\caption{
Period evolution of the $\ell$\,=\,1 (purple) and $\ell$\,=\,2 (green) g-modes at radial orders $n$=1,2,3,10
as the baseline model cools. Each point represents a timestep in \MESA\ where the g-mode was calculated by \GYRE.
The gray band show the \Teff\ range of observed DBV WD.} 
\label{fig:gyre_evol}
\end{figure}

Figure \ref{fig:N2equation1} compares the approximation in 
Equation (\ref{eq:bv_simple}) with the full $N^2$ calculation from \MESA.
The difference at 
$r/R$\,$\simeq$\,0.5 corresponds to the $X$(\oxygen[16]) $\rightarrow$ $X$(\carbon[12]) transition,
at $r/R$\,$\simeq$\,0.8 to the \carbon[12] bump, and
at $r/R$\,$\simeq$\,0.9 to the transition to
a He dominated atmosphere. 
Except for the outermost layers
and regions where the composition gradients
are significant, the agreement is sufficient 
to use Equation (\ref{eq:bv_simple}) as a scaling relation for building physical insights.
We always use, however, the full $N^2$ calculation from \MESA\ for any quantitative analysis. 

It is useful to reference features of the baseline model with respect to mass
or radius.  Figure \ref{fig:mr} thus shows the mass-radius relation of the
baseline model at $\log(L/\Lsun)$\,=\,$-$2.5 with key transitions labeled.

\begin{figure}[!htb]
\centering
\includegraphics[trim={1.0cm 1.5cm 0.0cm 1.75cm},clip,width=1.0\apjcolwidth]{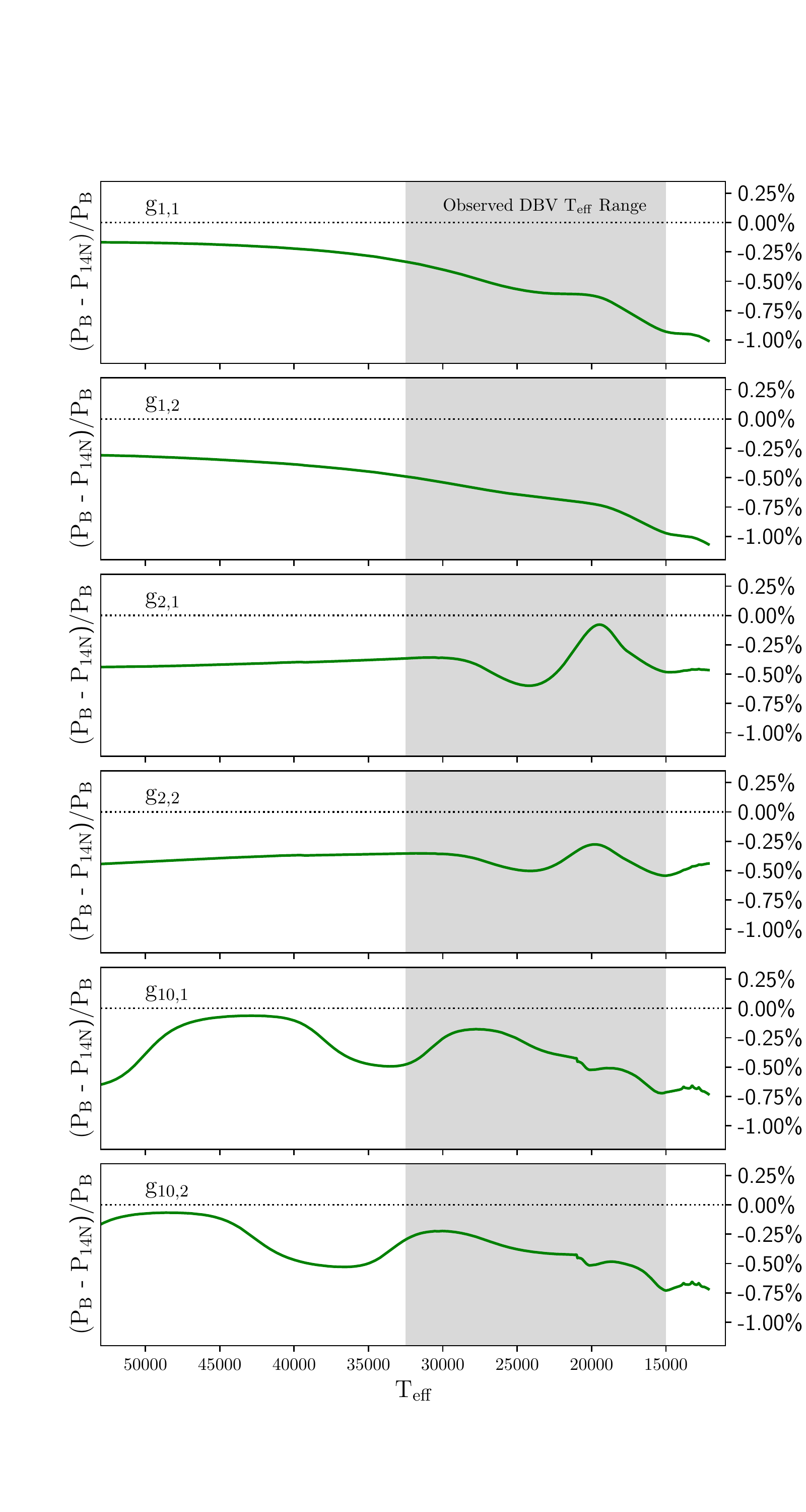}
\caption{{\it Top to Bottom:} Relative period differences of the 
$g_{1,1}$, $g_{1,2}$, $g_{2,1}$, $g_{2,2}$, $g_{10,1}$ and $g_{10,2}$ modes 
between the baseline model, P$_B$, and a model where the \neon[22] has
been replaced with \nitrogen[14], P$_{14\rm{N}}$. 
We use the notation $g_{n,\ell}$ for a g-mode of order $n$ and degree $\ell$.  
Gray bands show the $\Teff$ range of currently observed DBV WDs.
}
\label{fig:pdiffs_n14}
\end{figure}

\begin{figure}[!htb]
\centering
\includegraphics[trim={1.0cm 2.0cm 0.0cm 2.5cm},clip,width=1.0\apjcolwidth]{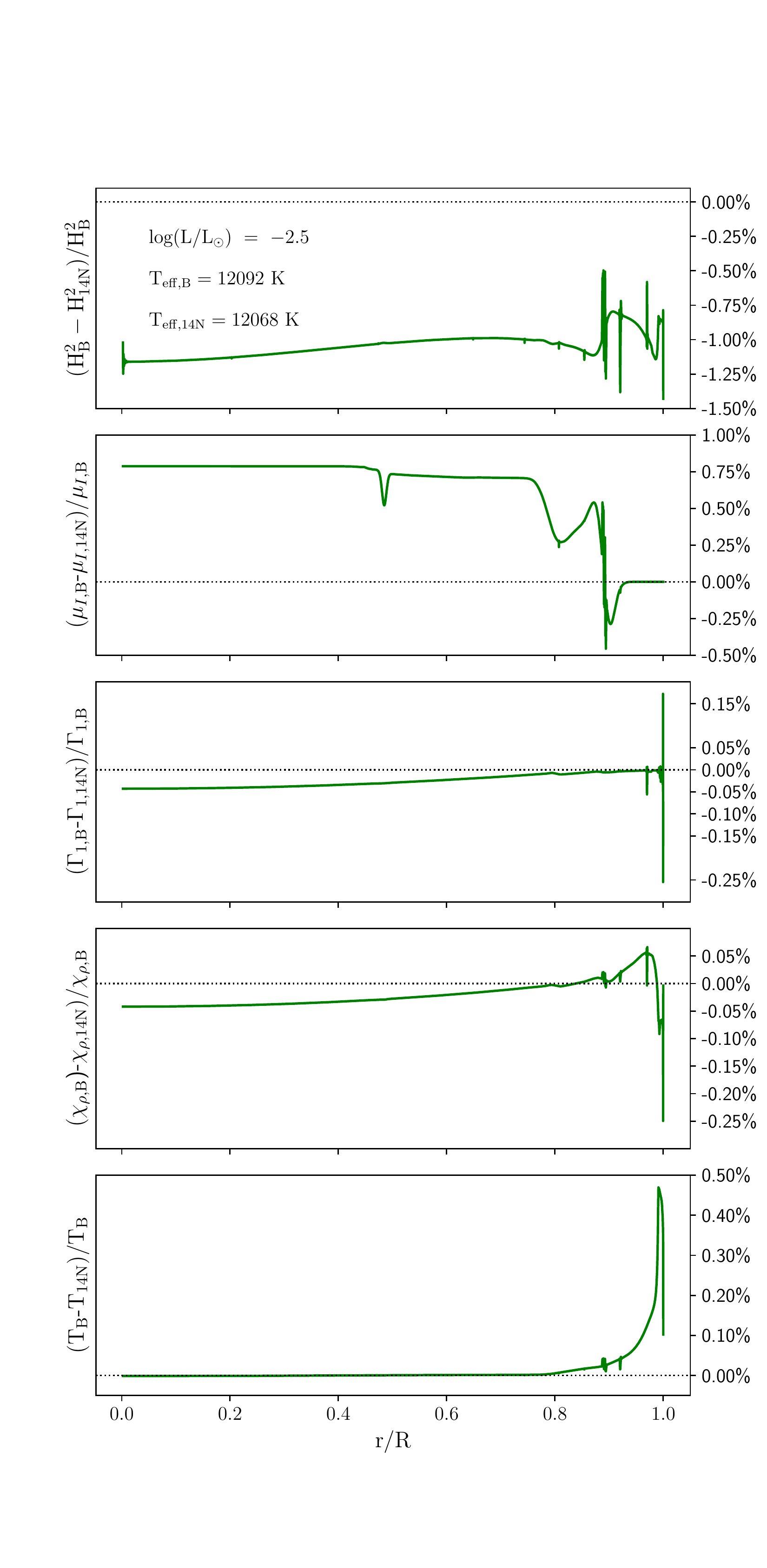}
\caption{{\it Top to Bottom:} Relative differences in the 
$H^2$, $\mu_I$ $\Gamma_1$, $\chi_{\rho}$, and $T$ 
contributions to $N^2$ in Equation~\ref{eq:bv_simple}.
Subscript B represents the baseline model, and subscript 14N represents
and a model where \neon[22] has been replaced with \nitrogen[14].
}
\label{fig:why_pdiffs_n14}
\end{figure}

Figure~\ref{fig:gyre_evol} shows the low-order g-mode pulsation periods
as the baseline WD model cools.  The periods
increase due to $N^2$ decreasing as the cooling
progresses, per Equation~\ref{eq:bv_simple}.  Higher radial orders
have steeper slopes due to the periods scaling with $k_r$ in
Equation~\ref{eq:gmodefrequency}. 
The increase in number of \MESA\ models at $\Teff$\,$\simeq$\,30,000\,K is due to 
the partial ionization of He, which leads to envelope convection in relatively hot DBV WDs.
The change in slope at $\Teff$\,$\simeq$\,20,000\,K is due to 
the luminosity becoming proportional to the enclosed mass, $L_r \propto M_r$,
as the plasmon neutrino emission becomes insignificant.

In Appendix~\ref{sec:convergence} we show that the low-order g-mode pulsation periods of the
baseline model calculated with \GYRE\ are only weakly dependent on the
mass and temporal resolution of the \MESA\ calculations.

\begin{figure*}[!htb]
\centering
\includegraphics[trim={0.5cm 0.7cm 2.0cm 0.5cm},clip,width=1.6\apjcolwidth]{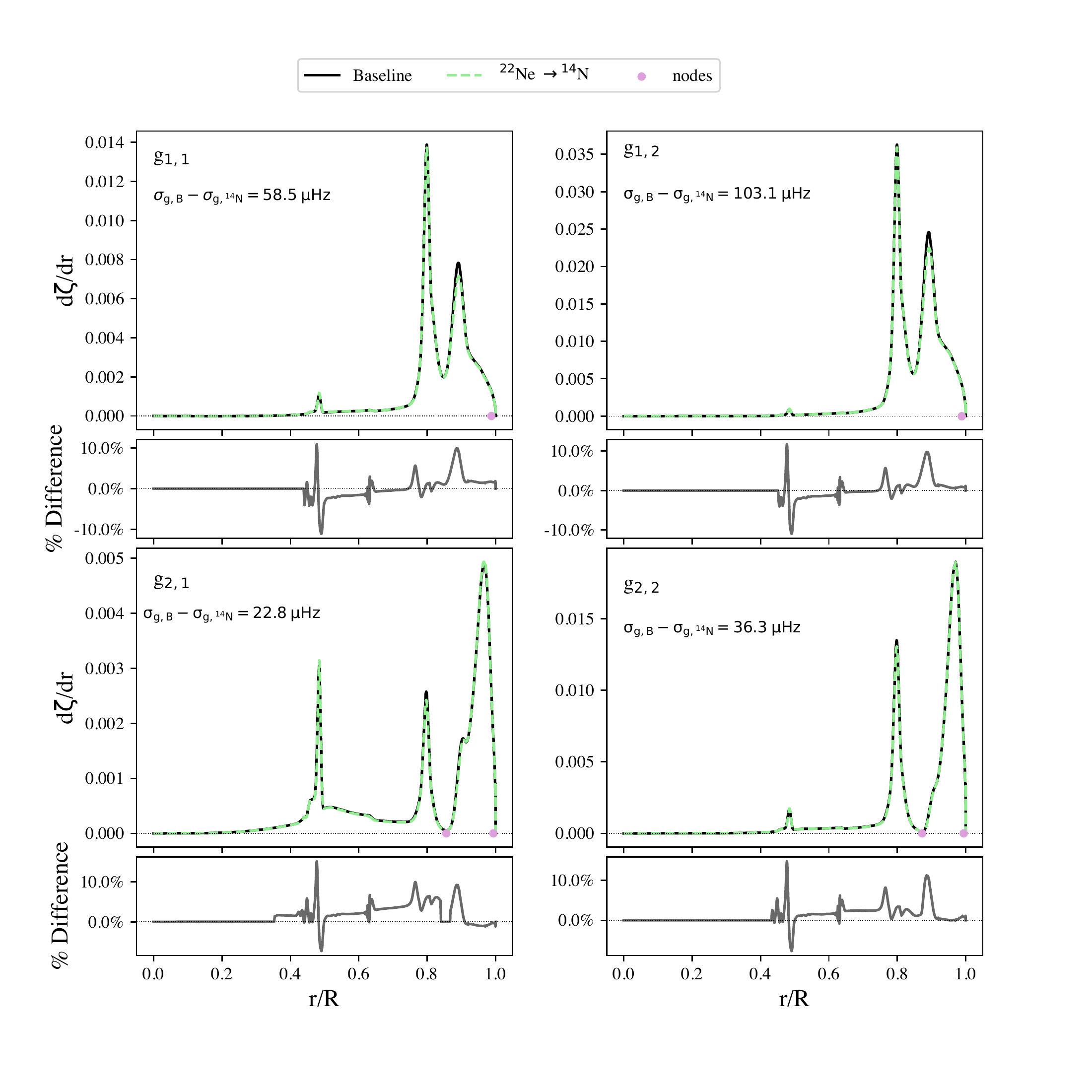}
\caption{Weight functions of the low-order g-modes for baseline model with \neon[22] (black curves) 
and a baseline model where \neon[22] has been replaced with \nitrogen[14] (green curves).
Subpanels show the relative percent differences between the two curves.
The profiles shown are when the two models have cooled to $\log(L/\Lsun)$\,=\,$-$2.5.
Nodes in the radial eigenfunctions are marked by filled circles.
}
\label{fig:n14_weights}
\end{figure*}

\begin{figure}[!htb]
\centering
\includegraphics[trim={0.0cm 0.0cm 0.0cm 0.0cm},clip,width = 1.0\apjcolwidth]{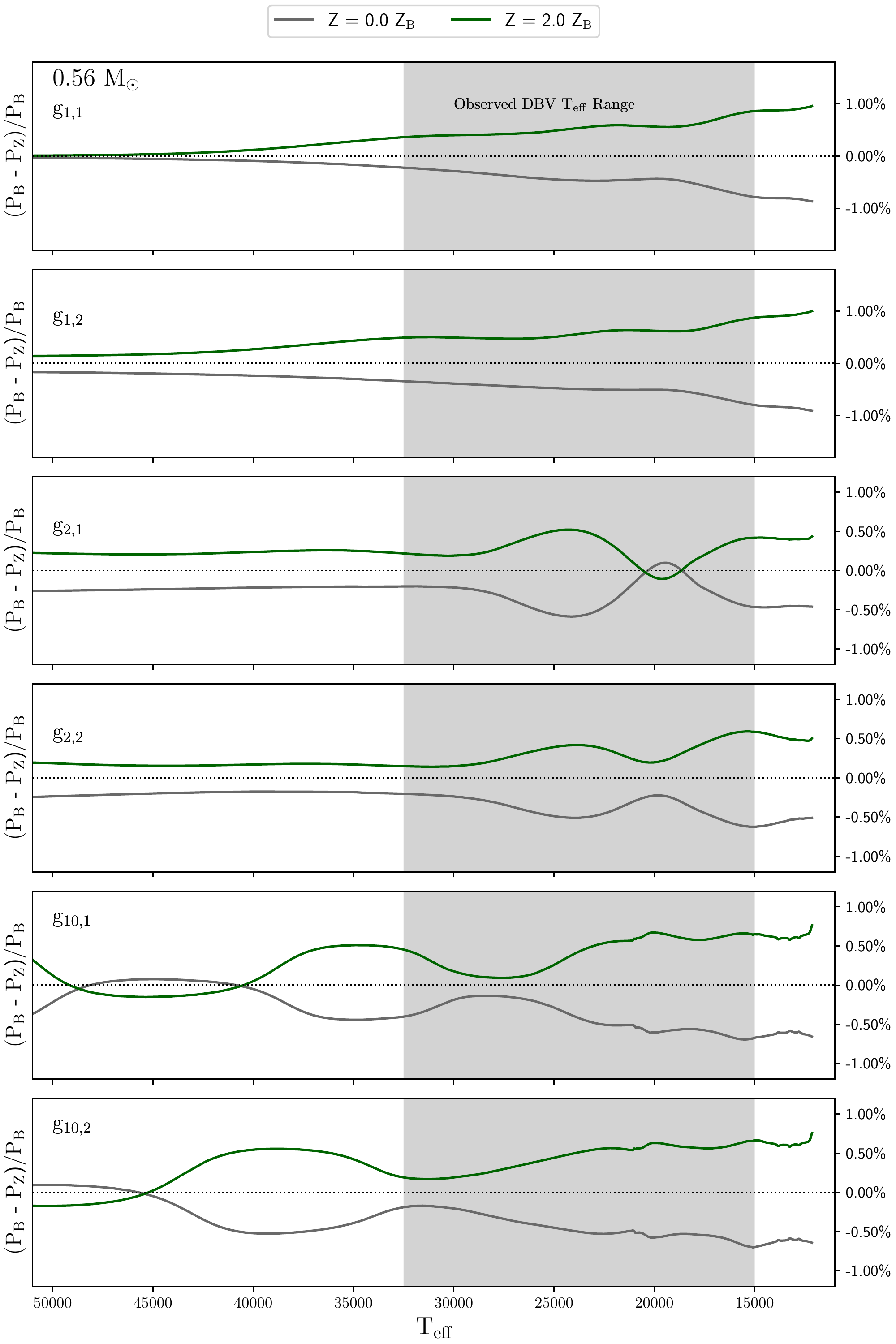}
\caption{{\it Top to Bottom:} Relative period differences of the $g_{1,1}$, $g_{1,2}$, $g_{2,1}$, $g_{2,2}$,
$g_{10,1}$ and $g_{10,2}$  modes 
between the baseline model, P$_B$, a zero-metallicity WD model (gray curves) where the \nitrogen[14] and \neon[22] 
have been put into \helium[4] and \carbon[12] respectively, and a super-solar metallicity model (green curves)
where the \nitrogen[14] and \neon[22] of the baseline model are doubled.
}
\label{fig:simp2_dperiods}
\end{figure}

\begin{figure}[!htb]
\centering
\includegraphics[trim={1.0cm 2.45cm 0.0cm 2.5cm},clip,width=1.0\apjcolwidth]{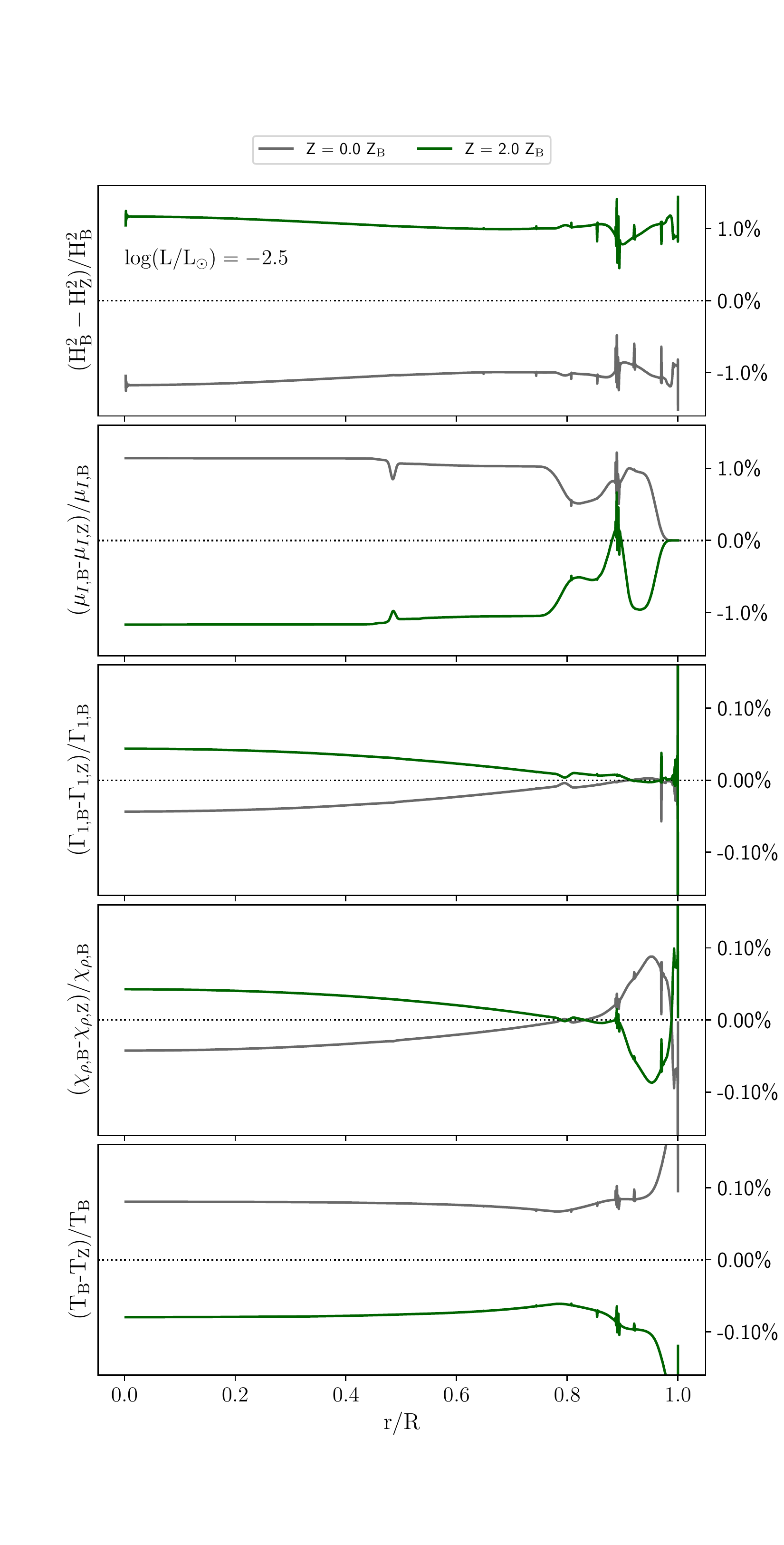}
\caption{{\it Top to Bottom:} Relative differences in the
$H^2$, $\mu_I$ $\Gamma_1$, $\chi_{\rho}$, and $T$
contributions to $N^2$ in Equation~\ref{eq:bv_simple}.
Subscript B represents the baseline model, and subscript Z represents
the zero-metallicity models (gray curves) and super-solar metallicity models (green curves).
}
\label{fig:simp2_eq2-parts}
\end{figure}

\vspace{0.3in}
\section{The Impact Of \neon[22]}\label{s.pchanges}

Having established the cooling properties and g-mode pulsation periods of a
baseline model whose mass fraction profiles are from a stellar
evolution model, we now explore changes in the g-mode pulsation periods due
to changes in the \neon[22] mass fraction profile shown in Figure~\ref{fig:illus}.
We consider three modifications: replacing \neon[22] with \nitrogen[14], 
a metal-free model, and a super-solar metallicity model.

\subsection{Putting the $^{22}$Ne into $^{14}$N}\label{s.pchanges.n14}

Replacing $X$(\neon[22]) with $X$(\nitrogen[14]) is a model for the reaction sequence 
\nitrogen[14]($\alpha$,$\gamma$)$^{18}$F(,$e^{+}\nu_e$)$^{18}$O($\alpha$,$\gamma$)\neon[22]
either physically not occurring or being ignored. 
Figure~\ref{fig:pdiffs_n14} shows the relative differences in the low-order g-mode pulsation periods
from this composition change.
All of the relative differences are negative, implying the pulsation periods in models that exclude \neon[22]
are longer than the corresponding pulsation periods in models that include \neon[22].
The magnitude of the relative period differences span $\simeq$\,0.25\%$-$1\% over the range of currently observed DBV WDs,
with the $g_{1,1}$ and $g_{1,2}$ modes showing the largest differences at cooler \Teff.
The change in the slopes at $\Teff$\,$\simeq$\,20,000\,K is due to 
plasmon neutrino emission becoming insignificant, and thus 
the luminosity becoming proportional to the enclosed mass, $L_r \propto M_r$.

What drives these g-mode period changes?
Replacing an isotope which has a larger mass number with an isotope which
has a smaller mass number decreases $\mu_I$. This replacement also increases $H$
through the mechanical structure and equation of state of the CO WD.  
Figure~\ref{fig:why_pdiffs_n14}
shows the relative differences in the $H^2$, $\mu_I$, $\Gamma_1$, $\chi_{\rho}$ and $T$ 
contributions to $N^2$ in Equation~\ref{eq:bv_simple}.
These changes collectively determine the magnitude and sign of the period change relative to the baseline model.
For this $X$(\neon[22]) $\rightarrow$ $X$(\nitrogen[14]) model,
the overall positive changes in $\mu_I$ and $T$ 
are counteracted by the negative changes from $H^2$, $\Gamma_1$, and $\chi_{\rho}$.
The magnitude of the relative difference in $H^2$ drives the net result
of a smaller $N^2$ and thus longer g-mode periods.
The nearly uniform negative change in $H^2$ imply a change in the radius of the WD model.
We find $(R_B - R_{14N})/R_B$\,$\simeq$\,$-$0.4\%, meaning  
the $X$(\neon[22]) $\rightarrow$ $X$(\nitrogen[14]) model
has a larger radius than the model with \neon[22].
This is expected given differences in the electron fraction of a WD.

Figure \ref{fig:n14_weights} compares the weight functions 
of the baseline model with \neon[22] and the model where the 
\neon[22] has been replaced with \nitrogen[14].
Following \citet{kawaler_1985_aa}, the  weight function is

 \begin{equation}
  \frac{{\rm d}\zeta}{{\rm d}r} = \frac{[C({\bf y},r) + N({\bf y},r) + G({\bf y},r)] \rho r^2 }{\int_{r=0}^{r=R} T({\bf y},r) \rho r^2 {\rm d}r}
\ ,
\label{eq:dzdr}
\end{equation}
where 
$C({\bf y},r)$ varies with the Lamb frequency,
$N({\bf y},r)$ contains the \BV\ frequency,
$G({\bf y},r)$ involves the gravitational eigenfunctions,
$T({\bf y},r)$ is proportional to the kinetic energy density,
and ${\bf y} = (y_1,y_2,y_3,y_4)$ are the \citet{dziembowski_1971_aa} variables.
The frequency of an adiabatic mode is then 

\begin{equation}
\nu^2 =  \zeta = \int_{r=0}^{r=R} \frac{{\rm d}\zeta}{{\rm d}r} \cdot {\rm d}r
\ .
\label{eq:work}
\end{equation}

The weight function for the two models is dominated by 
the $N({\bf y},r)$ term except for the surface layers.
Figure \ref{fig:n14_weights} shows that the net effect of the 
\neon[22] $\rightarrow$ \nitrogen[14] composition
change is a shift in $\zeta$, the area under the weight function curves, towards
smaller frequencies of the low-order g-modes. 
The subpanels in Figure \ref{fig:n14_weights} illustrate the relative percent differences between the weight function curves.
Most of the changes in $\zeta$ occur at the CO transition region 
($r/R$\,$\simeq$\,0.45, see Figure~\ref{fig:mr}), \carbon[12] bump ($r/R$\,$\simeq$\,0.8
), and at the transition
to a He-dominated atmosphere ($r/R$\,$\simeq$\,0.9).  The changes in these regions get as large as $\sim$ 10\%.
We identify the dipole g-mode of radial order $n$\,=\,2 as being
more sensitive to the location and gradient of $\mu_I$ at the CO
transition ($r/R$\,$\simeq$\,0.5) than other low-order g-modes.

\subsection{Zero-Metallicity and Super-Solar Metallicity}\label{s.pchanges.usual}

Replacing $X$(\nitrogen[14]) with $X$(\helium[4]) and $X$(\neon[22]) with $X$(\carbon[12])
is a model for ignoring the birth metallicity of the ZAMS star, CNO burning on the
main-sequence, and the \nitrogen[14]($\alpha$,$\gamma$)$^{18}$F(,$e^{+}\nu_e$)$^{18}$O($\alpha$,$\gamma$)\neon[22]
reaction sequence during He-burning.  Most studies of the pulsation periods
of observed WDs use zero-metallicity DBV WDs when deriving the interior mass fraction profiles,
although see \citet{camisassa_2016_aa} for a counterexample.
Alternatively, doubling $X$(\nitrogen[14]) at the expense of $X$(\helium[4])
and doubling $X$(\neon[22]) at the expense of $X$(\carbon[12]) is a model
for a super-solar metallicity DBV WD.

Figure~\ref{fig:simp2_dperiods} compares the relative change in the
low-order g-mode pulsation periods of the zero and
super-solar metallicity models.  The period differences are negative
for the zero-metallicity model and positive for the super-solar
metallicity model.  Zero-metallicity DBV WD models have longer periods than
the baseline model, which in turn has longer periods than the super-solar
metallicity model.  The relative period differences 
of the zero and super-solar metallicity models 
are mostly symmetric about
the baseline model's $Z$\,=\,0.02 metallicity.  
The period differences of the zero-metallicity models, 
averaged over the $\Teff$ evolution, are 
$\Delta P(g_{1,1})$\,$\simeq$\,$-$0.57\,s,
$\Delta P(g_{1,2})$\,$\simeq$\,$-$0.40\,s,
$\Delta P(g_{2,1})$\,$\simeq$\,$-$0.52\,s, and
$\Delta P(g_{2,2})$\,$\simeq$\,$-$0.40\,s.
For the super-solar metallicity models the  averaged absolute period differences are
$\Delta P(g_{1,1})$\,$\simeq$\,0.66\,s,
$\Delta P(g_{1,2})$\,$\simeq$\,0.45\,s,
$\Delta P(g_{2,1})$\,$\simeq$\,0.46\,s, and
$\Delta P(g_{2,2})$\,$\simeq$\,0.35\,s.
Over the $\Teff$ range of currently observed DBV WDs, 
the mean relative period change of the dipole modes is 0.57\% and the maximum of relative period change is 0.88\%.
The relative period change of the quadrupole  modes is smaller, with a mean of 0.33\% and
a maximum of 0.63\%.

Figure~\ref{fig:simp2_eq2-parts} shows the relative differences  
in the $H^2$, $\mu_I$, $\Gamma_1$, $\chi_{\rho}$ and $T$ 
contributions to $N^2$ of Equation~\ref{eq:bv_simple}
for the zero and super-solar metallicity models.
These changes collectively determine the magnitude and sign of the period change relative to the baseline model.
For the zero-metallicity models, the combined positive changes in $\mu_I$ and $T$ 
are counteracted by the collective negative changes from  $H^2$, $\Gamma_1$, and $\chi_{\rho}$.
The net change is negative, resulting in smaller $N^2$ and longer g-mode periods.
Similar reasoning for the super-solar metallicity models leads to a net 
positive change, resulting in larger $N^2$ and smaller g-mode periods.
The magnitude of the difference in $H^2$ drives the overall result
for both metallicity cases.
The nearly uniform changes in $H^2$ imply changes in the radii,
and we find $(R_B - R_Z)/R_B$\,$\simeq$\,$\pm$0.4\% with zero-metallicity models
having smaller radii and super-solar metallicity models having larger radii.

\begin{figure*}[!htb]
\centering
\includegraphics[trim={0.0cm 0.0cm .0cm 0.0cm},clip,width=1.0\textwidth]{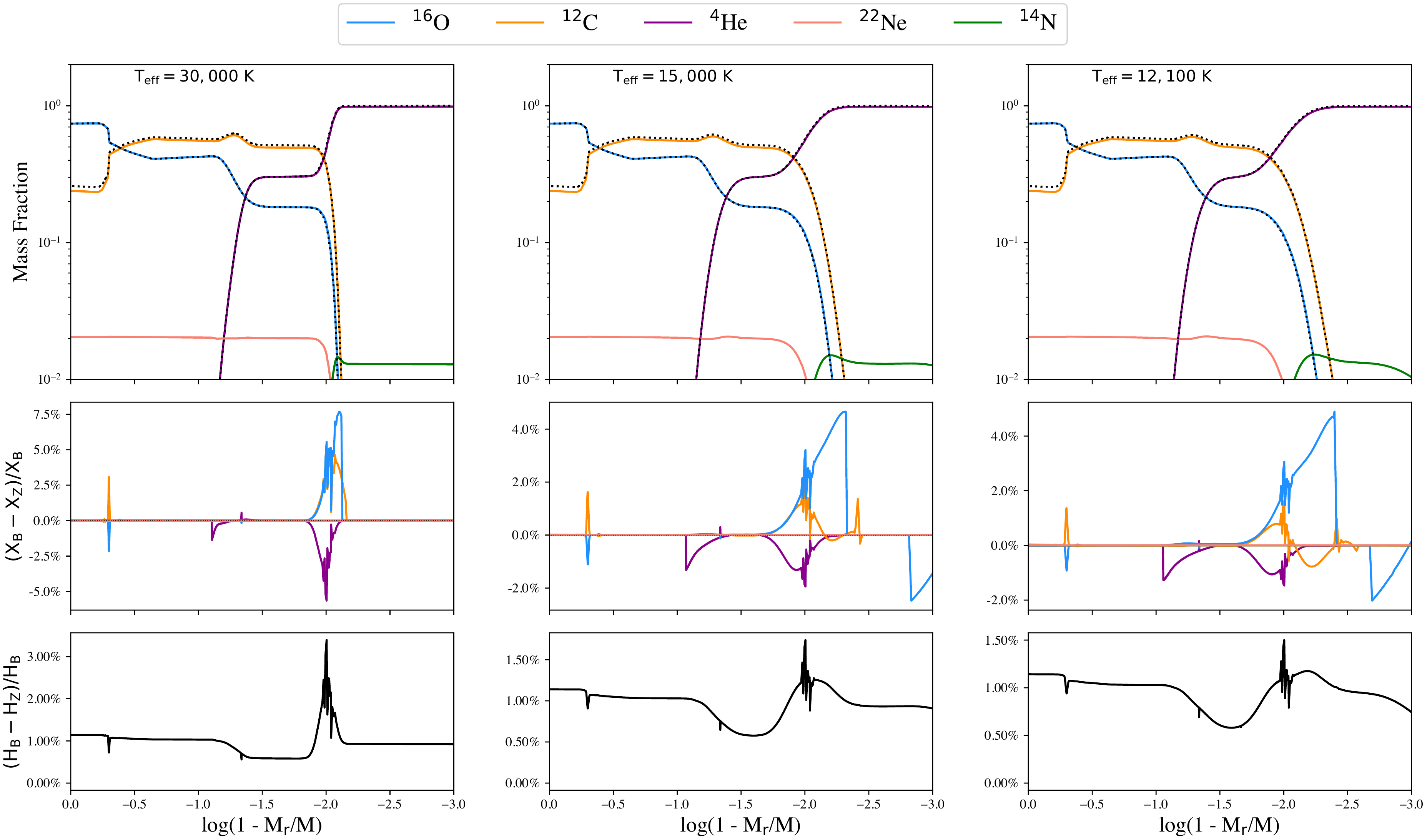}
\caption{\textit{Top Panels:} Mass fraction profiles for 0.56\,\Msun baseline (colored curves)  
and zero metallicity (black dashed curves) models at \Teff\,$\simeq$\,30,000~K, 15,000~K, and 12,100~K. 
\textit{Middle Panels:} Relative differences in mass fraction profiles, 
where we have zeroed out the \neon[22] and \nitrogen[14] 
offsets from \carbon[12] and \helium[4] respectively.  
\textit{Bottom Panel}: Relative differences in $\mu_I$. }
\label{fig:compare_baseline_zero}
\end{figure*}

Interrogating further the composition dependence, 
the top panels of Figure~\ref{fig:compare_baseline_zero} compare the mass fraction profiles 
of the $X$(\neon[22])\,$\simeq$\,0.02 baseline and zero-metallicity at 30,000\,K, 15,000\,K and 12,100\,K as a function
of mass coordinate. Element diffusion is operating in both models. The middle panels show the relative differences 
in these mass fraction profiles, with the \neon[22] and \nitrogen[14] offsets zeroed out.
The C and O differences at $\log(1 - M_r/M)$\,$\simeq$\,$-$0.25, from Figure~\ref{fig:mr}, correspond to the C/O transition at $r/R$\,$\simeq$\,0.5.
The He difference at $\log(1 - M_r/M)$\,$\simeq$\,$-$1.0 correlates to the rise of He at $r/R$\,$\simeq$\,0.75.
Similarily, the C, O and He differences at $\log(1 - M_r/M)$\,$\simeq$\,$-$2.0 maps to He dominating the composition at $r/R$\,$\simeq$\,0.9.
These relative differences are the largest at
30,000\,K, reaching $\simeq$\,7.5\% for \oxygen[16] and $\simeq$\,$-$6\% for \helium[4]. 
The relative differences at 15,000\,K and 12,100\,K have about the same magnitude,
$\simeq$\,7.5\% for \oxygen[16 ] and $\simeq$\,$-$1\% for \helium[4].
The relative mass fraction differences span a larger range of $\log(1 - M_r/M)$ 
as the models cool due to element diffusion.
The bottom panels of Figure~\ref{fig:compare_baseline_zero} show the corresponding 
relative difference in the $\mu_{I}$ profiles.
As $\mu_{I}$ is calculated by dividing the mass fraction of a isotope by its atomic weight, 
the relative differences in the mass fraction profiles are reduced in the $\mu_{I}$ profiles.
The $\mu_{I}$ profile for 12,100\,K in terms of a mass coordinate is the same 
as the $\mu_{I}$ profile in Figure~\ref{fig:simp2_eq2-parts} in terms of a radial coordinate.

\begin{figure}[!htb]
\centering
\includegraphics[trim={1.2cm 2.2cm .0cm 0.0cm},clip,width=1.0\apjcolwidth]{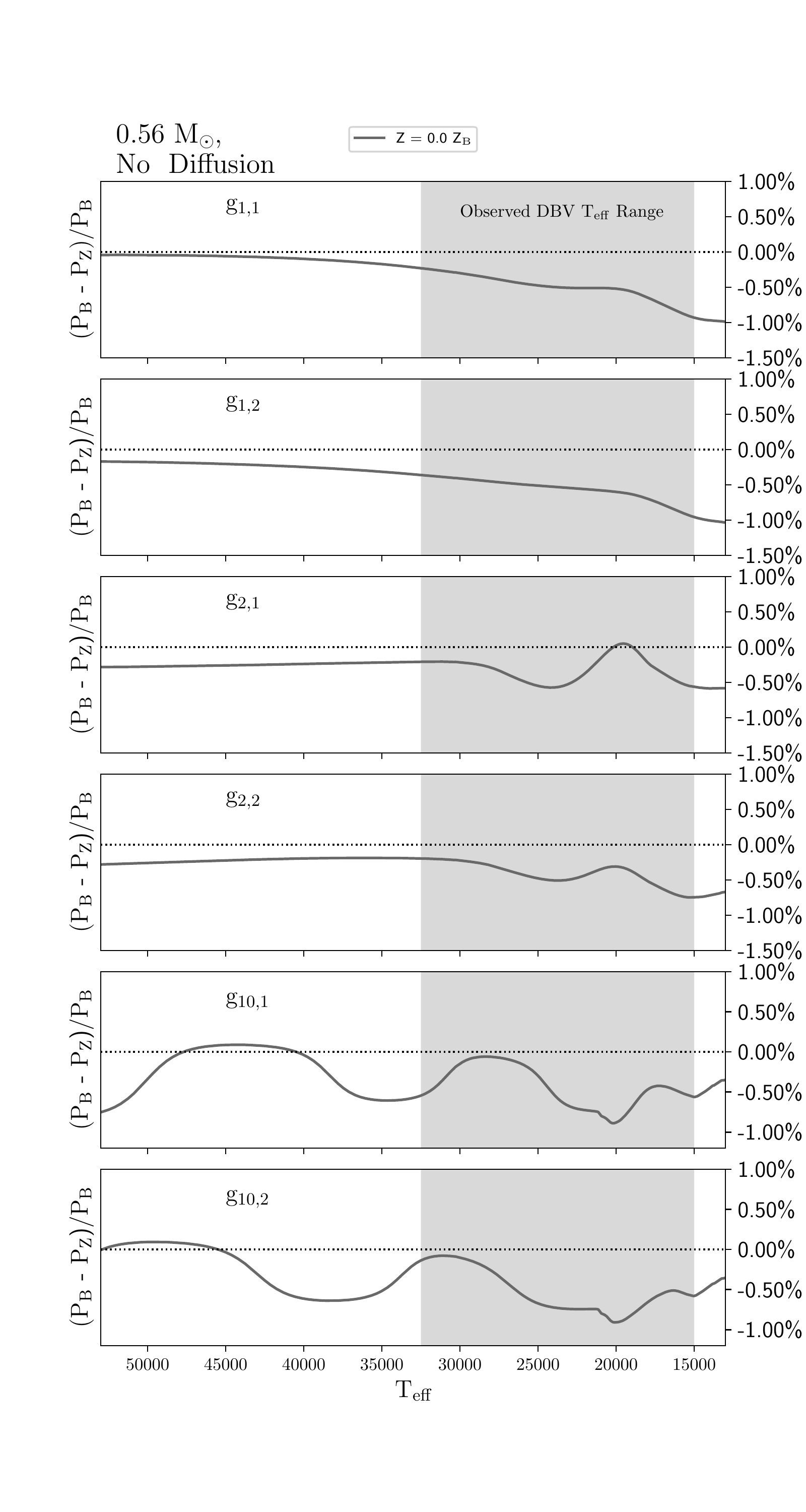}
\caption{{Top to Bottom: Relative period differences of the $g_{1,1}$, $g_{1,2}$, $g_{2,1}$, $g_{2,2}$,
$g_{10,1}$ and $g_{10,2}$  modes 
between the baseline model, P$_B$, and the zero-metallicity WD model, P$_Z$, with diffusion turned off.}}
\label{fig:noDiffusion_base_zero}
\end{figure}

We also computed the relative period differences between the $X$(\neon[22])\,$\simeq$\,0.02 baseline 
and zero-metallicity model with diffusion turned off to disentangle structural and diffusion effects.
The results are shown in Figure~\ref{fig:noDiffusion_base_zero}.  
While there is a slight variation from the zero-metallicity gray curves shown in 
Figure~\ref{fig:simp2_dperiods}, mostly in the higher order $g_{10,1}$ and $g_{10,2}$ modes, 
the magnitude of the relative differences remains the same.  
This further suggests that the period shifts are a direct consequence 
of the presence or absence of \neon[22].

\begin{figure*}[!htb]
\centering
\includegraphics[trim={0.0cm 3.0cm .0cm 3.5cm},clip,width=.9\apjcolwidth]{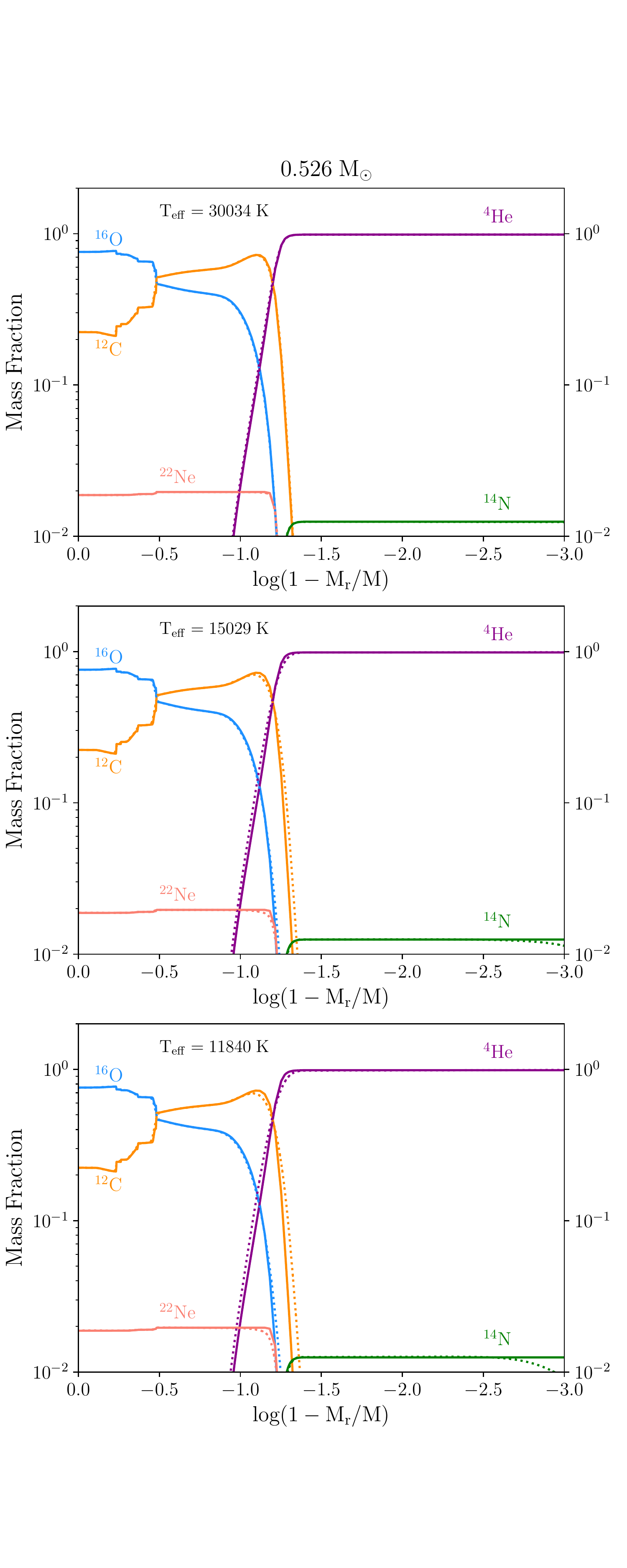}
\includegraphics[trim={0.0cm 3.0cm .0cm 3.5cm},clip,width=.9\apjcolwidth]{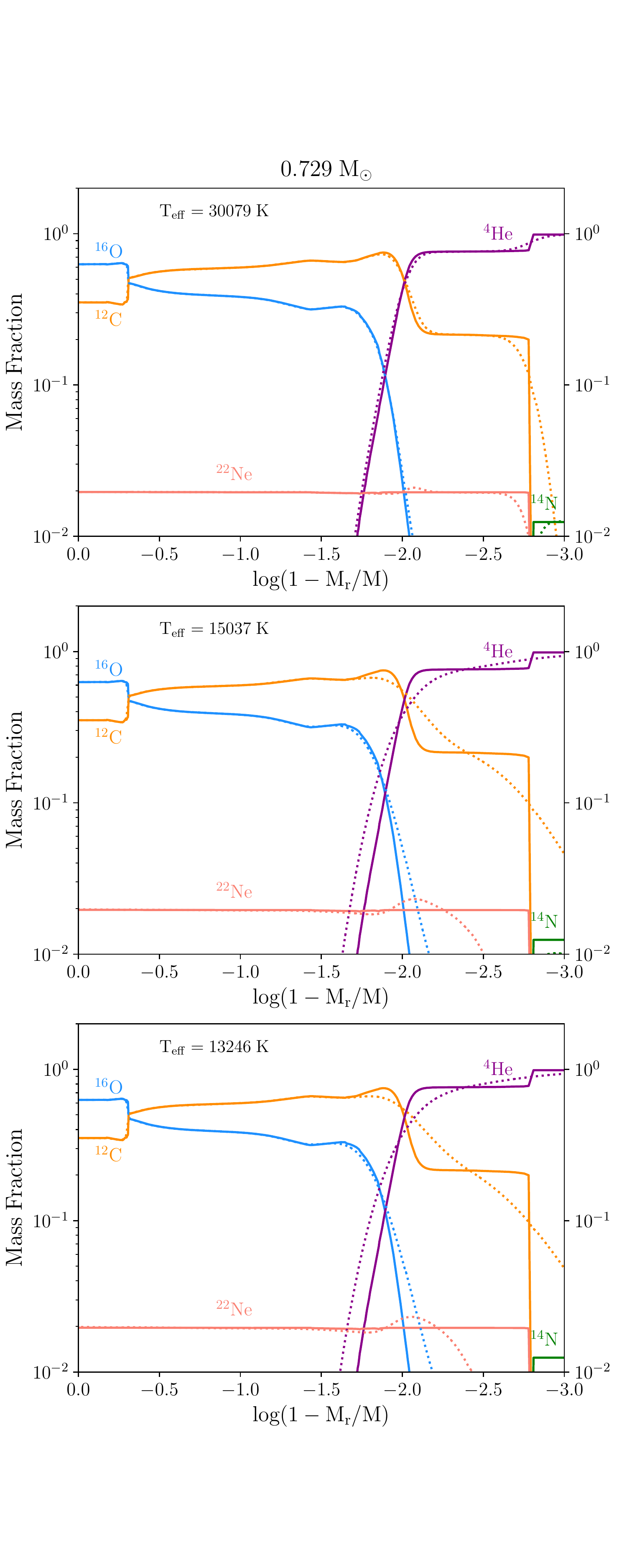}
\caption{Mass fractions profiles for 0.52\,\Msun (left column) and 0.73\,\Msun (right column) 
ab initio DB WD models at \Teff\,$\simeq$\,30,000~K (top), 15,000~K (middle),
and at the end of the evolution (bottom). Initial mass fraction profiles are shown as solid curves 
and the diffusing mass fraction profiles are shown as dotted curves.}
\label{fig:0.526_mfracs}
\end{figure*}

\begin{figure*}[!htb]
\centering
\makebox[\apjcolwidth]{\includegraphics[trim={4.cm 2.5cm 1cm 0.cm},clip,width = 1.0\textwidth]{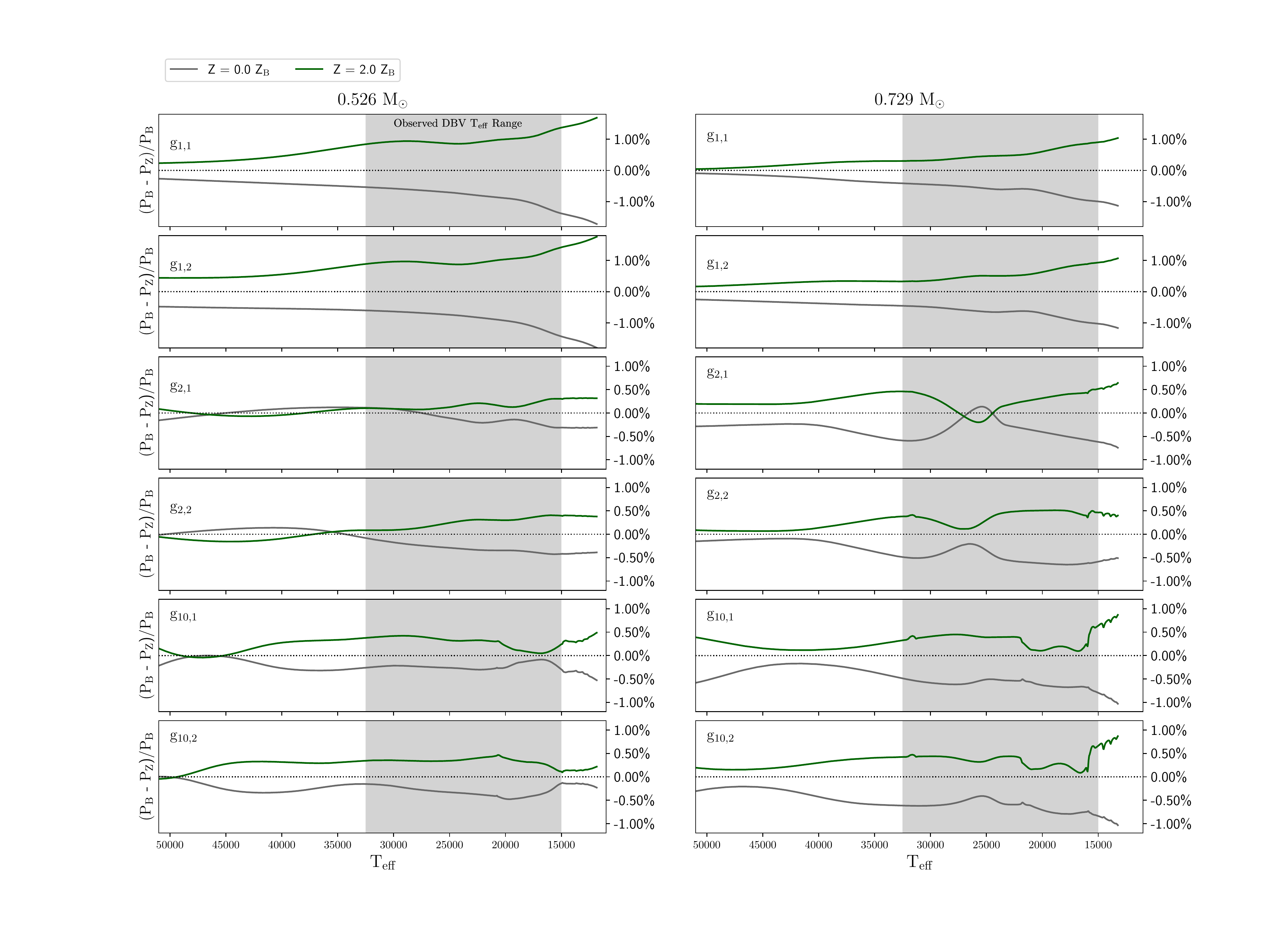}}
\caption{Relative period differences of the $g_{1,1}$, $g_{1,2}$, $g_{2,1}$, $g_{2,2}$, $g_{10,1}$ and $g_{10,2}$ modes 
for 0.526\,\Msun (left column) and 0.729\,\Msun (right column). Differences are 
between the baseline model, P$_B$, a zero-metallicity WD model (gray curves) where the \nitrogen[14] and \neon[22] 
have been put into \helium[4] and \carbon[12] respectively, and a super-solar metallicity model (green curves)
where the \nitrogen[14] and \neon[22] of the baseline model are doubled.
}
\label{fig:dperiods_p53_p73}
\end{figure*}

\section{Trends in the period changes with the white dwarf mass}\label{s.others}

Using the same physics and numerical choices as for the 0.56\,\Msun\ baseline model,
we evolved a $Z$\,=\,0.02, 1.1\,\Msun\ ZAMS stellar model from the pre-main sequence to a 0.526\,\Msun\ DB WD,
and a $Z$\,=\,0.02, 3.6\,\Msun\ ZAMS model to a 0.729\,\Msun\ DB WD. 
This initial to final mass mapping is similar to Table 1 of \citet{camisassa_2016_aa}.
Relative to the 0.56\,\Msun\ baseline model, the 0.526\,\Msun\ WD model has a thicker He-layer
and a more abbreviated extent of $X$(\neon[22]). Conversely, the  0.729\,\Msun\ WD model has a 
smaller \carbon[12] bump, a thinner He-layer, and a more extended  $X$(\neon[22]) profile.
These mass fraction profiles were imposed on 0.526\,\Msun and 0.729\,\Msun\ ab initio WD models,
respectively. 

Figure \ref{fig:0.526_mfracs} shows the diffusion of these initial mass fraction
profiles as the ab initio WD models cool to \Teff\,$\simeq$\,30,000~K, then $\simeq$\,15,000~K and finally 
$\simeq$\,12,000~K (corresponding to the termination at $\log(L/\Lsun)$\,=\,$-$2.5).
Element diffusion is more pronounced for the more massive 0.729\,\Msun\ DB WD model due to its
larger surface gravity. An enhancement forms
in the $X$(\neon[22]) profile at $\log(1 - M_r/M$)\,$\simeq$\,$-2.0$ by the time the
0.729\,\Msun\ model has cooled to \Teff\,$\simeq$\,30,000~K.
As the model further cools, the $X$(\neon[22]) bump grows in amplitude 
as it propagates inwards toward the center through the He-dominated outer layers.
The $X$(\neon[22]) bump generates an increase in the local $N^2$ in the regions it propagates through
from a locally larger $\mu_I$ and a smaller compensating $H^2$. The regions trailing the  $X$(\neon[22]) bump
are depleted of $X$(\neon[22]), causing a decrease in the local $N^2$ in these regions.

We find longer low-order g-mode periods for the more massive WD, consistent with \citet{camisassa_2016_aa}.
As was done for the 0.56\,\Msun\ baseline model, we replace 
$X$(\nitrogen[14]) with $X$(\helium[4]) and $X$(\neon[22]) with $X$(\carbon[12])
to generate a zero-metallicity ab initio DB WD  model.
We also double $X$(\nitrogen[14]) at the expense of $X$(\helium[4])
and double $X$(\neon[22]) at the expense of $X$(\carbon[12]) to generate 
a super-solar metallicity DB WD.

Figure~\ref{fig:dperiods_p53_p73} compares the relative change in the
low-order g-mode pulsation periods of the zero and
super-solar metallicity 0.526\,\Msun\ and  0.729\,\Msun\ DB WD models.
As for the 0.56\,\Msun\ baseline model, the relative period differences 
are mostly symmetric about
the reference model's $Z$\,=\,0.02 metallicity.  
For the 0.526\,\Msun\ models, over the $\Teff$ range of currently observed DBV WDs,
the mean relative period change of the dipole modes is 0.99\% and the maximum of relative period change is 1.43\%.
The relative period change of the quadrupole  modes is smaller, with a mean of 0.25\% and
a maximum of 0.43\%.
For the 0.729\,\Msun\ models,
the mean relative period change of the dipole modes is 0.65\% and the maximum of relative period change is 1.02\%.
The relative period change of the quadrupole  modes is again smaller, with a mean of 0.40\% and
a maximum of 0.65\%.
These values are commensurate with the mean and maximum relative period changes found for
the  0.56\,\Msun\ baseline model.

There are a few trends in the relative period differences with respect to the WD mass. 
For the zero-metallicity $n$\,=\,2 and $n$\,=\,10 g-modes, the average relative differences 
in the observed T$\rm_{eff}$ range increase with increasing mass.  For example, as the WD mass is 
increased from 0.526\,M$_\odot$ to 0.560\,M$_\odot$, we find 
the average relative period differences increase by factors of 
1.74, 1.22, 2.43, and 1.46, for the g$_{2,1}$, g$_{2,2}$, g$_{10,1}$, and g$_{10,2}$ modes, respectively.  
As the WD mass is further increased from 0.560\,M$_\odot$ to 0.729\,M$_\odot$, we find additional magnification 
factors  of 1.21, 1.29, 1.21, and 1.26, for g-modes g$_{2,1}$, g$_{2,2}$, g$_{10,1}$, and g$_{10,2}$ respectively.  
The absence of \neon[22] causes a greater deviation from the reference metallicity model as the WD mass increases.

The g$_{2,1}$ and g$_{2,2}$ g-modes show a trend in the local minimum 
as the WD mass increases. For the 0.526\,M$_\odot$ model, the g$_{2,1}$ g-mode 
has a local minimum at $\Teff$\,$\lessapprox$\,20,000~K.
For the 0.526\,M$_\odot$ baseline model, this local minimum crosses zero at $\Teff$\,$\simeq$\,20,000~K.
For the 0.729\,M$_\odot$ model, the local minimum is deeper and crosses zero at 
at $\Teff$\,$\simeq$\,25,000~K. These trends with mass are due to when energy lost by the cooling WD
is no longer dominated by neutrino cooling.

\vspace{0.3in}
\section{Discussion} \label{s.summary}

We explored changes in the adiabatic low-order g-mode pulsation periods of
0.526, 0.560, and 0.729\,\Msun\ DB WD models due to the
presence, absence, and enhancement of \neon[22] as the models cool through
the observed range of effective temperatures.
We found mean relative period shifts of $\Delta
P/P$\,$\simeq$\,$\pm$\,0.5\% for the low-order dipole and
quadrupole g-mode pulsations within the observed effective temperature
window, with a range of $\Delta P/P$ that depends on the
specific g-mode, mass fraction of \neon[22], effective temperature,
and mass of the WD model. Shifts in the pulsation periods due to the 
presence, absence, or enhancement of $X$(\neon[22]) mostly arise from a competition
between the pressure scale height and ion mean molecular weight.

Low-mass DAV WDs, the ZZ Ceti class of stars, have pulsation periods in the 100$-$1500 s range \citep[e.g.,][]{vincent_2020_aa}.
Comparing low-mass DAV WDs produced from stellar evolution models with and without diffusion
of \neon[22], \citet{camisassa_2016_aa} find that the \neon[22] sedimentation induces 
mean period differences of $\simeq$\,3 s, reaching maximum period differences of $\simeq$\,11 s. 
For the more massive DAV WD models, where sedimentation of \neon[22] is stronger, they find 
mean period differences of $\simeq$\,15 s between when diffusion is on and off, 
and a maximum period differences of $\simeq$\,47 s. Comparatively, our article focuses on DBV WD models,
the evolution of the pulsation periods as the DBV WD models cool, and the impact of \neon[22] being present,
absent, or enhanced in the WD interior.
Nevertheless, we conducted an experiment of turning element diffusion off in our 0.56\,\Msun\ baseline model.  
At $\rm \log(L/\Lsun)$\,=\,$-2.5$, we find an absolute mean difference for $n$\,=\,1 to $n$\,=\,11 of $\simeq$\,16~s, 
with a maximum period difference at $n$\,=\,9 of $\simeq$\,56~s.  
This maximum difference equates to a $\simeq$\,7\% relative difference between when diffusion is on and off.
Our period changes are slightly higher than those found 
in \cite{camisassa_2016_aa}'s 0.833 M$_\odot$ model, and
much larger than the differences found in their 0.576 M$_\odot$ model.
These differences could be a result of DAV versus DBV models, 
as DAV models have different cooling times than DBV models.
In addition, \cite{camisassa_2016_aa} computes their period differences at 
$\rm log(L/L_\odot) = -2.80$ and $\rm log(L/L_\odot) = -2.93$ for their 0.576 and 0.833 M$_\odot$ models, respectively.  
These are dimmer than the $\rm \log(L/\Lsun)$\,=\,$-2.5$ used for our calculations. 
Our maximum radial order is found up to 11 at this luminosity, 
while \cite{camisassa_2016_aa} uses more radial orders, with a maximum radial order of 50.

\citet{giammichele_2018_aa} compares the g-mode pulsation periods of a pure oxygen core surrounded by a pure helium envelope 
with those from an oxygen-dominated core with $X$(\neon[22])\,=\,0.02 surrounded by a pure helium envelope. 
They find including \neon[22] yields shorter periods, with mean period differences of $\simeq$\,0.1\%.
We find a mean period shift that is about 5 times larger in our 0.56\,\Msun baseline model.
This difference may be caused by the contrast in the composition of the models, which in turn
causes variances in the local mean molecular weight and pressure scale height scaling described by Equation~\ref{eq:bv_simple}.

Are 1\% or less period differences important?
The g-mode periods of observed DBV WD
are found from a Fourier analysis of the photometric light curves
and are typically given to 6$-$7 significant figures of precision.  
Usually zero-metallicity WD models (i.e., without
\neon[22]) are fit to the observed g-mode periods and other properties (e.g., $\Teff$, $\log g$).
The root-mean-square residuals to the $\simeq$\,150$-$400~s low-order g-mode periods 
are typically in the range $\sigma_{\rm rms}$\,$\lesssim$\,0.3~s \citep[e.g.,][]{bischoff-kim_ostensen_2014}, 
for a fit precision of $\sigma_{\rm rms}/ P$\,$\lesssim$\,0.3\%.
Our finding of a mean relative period shift of $\Delta P/ P$\,$\simeq$\,$\pm$\,0.5\% 
induced by including 
\neon[22] in WD models suggests a systematic offset may be
present in the fitting process of specific WDs when \neon[22] is absent.
As part of the fitting process involves adjusting the
composition profiles of the model WD, this study on the impact of \neon[22] can inform
inferences about the derived interior mass fraction profiles.
We encourage routinely including \neon[22] mass fraction profiles, informed by
stellar evolution models, to future generations of DBV WD model fitting processes.

The adiabatic low-order g-mode pulsation periods of our DB WD models depend upon
simplifying assumptions in the stellar evolution calculations (e.g.,
convective boundary layer mixing, shellular rotation), uncertainties
(e.g., mass loss history, stripping of the residual thin H layer,
thickness of the He-dominated atmosphere), and unknown inherent
systematics.  We hypothesize that these model dependencies
and systematics could mostly cancel when dividing one model result by
another model result, such as when calculating the relative 
period shifts $\Delta P/P$. We anticipate exploring a larger range of
models, similar in approach to \citet{fields_2016_aa}, to test this
conjecture in future studies.

\acknowledgements

The \MESA\ project is supported by the National Science Foundation (NSF) under the Software Infrastructure for Sustained Innovation
program grants (ACI-1663684, ACI-1663688, ACI-1663696).
This research was also supported by the NSF under grant PHY-1430152 for the Physics Frontier Center 
``Joint Institute for Nuclear Astrophysics - Center for the Evolution of the Elements'' (JINA-CEE).
A.T. is a Research Associate at the Belgian Scientific Research Fund (F.R.S-FNRS).
We acknowledge useful discussions at virtual Sky House 2020.
This research made extensive use of the SAO/NASA Astrophysics Data System (ADS).

\software{
\MESA\ \citep[][\url{http://mesa.sourceforge.net}]{paxton_2011_aa,paxton_2013_aa,paxton_2015_aa,paxton_2018_aa,paxton_2019_aa},
\texttt{MESASDK} 20190830 \citep{mesasdk_linux,mesasdk_macos},
\code{wd\_builder} \url{https://github.com/jschwab/wd_builder},
\GYRE\ \citep[][\url{ https://github.com/rhdtownsend/gyre}]{townsend_2013_aa,townsend_2018_aa},
Montreal White Dwarf Database \citep[][\url{http://www.montrealwhitedwarfdatabase.org}]{dufour_2017_aa}, 
\texttt{matplotlib} \citep{hunter_2007_aa}, and
\texttt{NumPy} \citep{der_walt_2011_aa}.
         }
         
\clearpage

\appendix

\section{Convergence Studies}
\label{sec:convergence}

In this appendix we demonstrate that the pulsation periods of the baseline model
are only weakly dependent on the details of the mass and temporal resolution of 
the \MESA\ + \GYRE\ calculations.

A \MESA\ parameter controlling the mass resolution is
\texttt{max\_dq}, the maximum fraction a model's mass in one
cell. That is, the minimum number of cells in a model is
$N_{\rm min\ cells}$\,=\,1/\texttt{max\_dq}.  
We use $N_{\rm min \ cells}$\,=\,5,000 for all the results reported.
\MESA\ can also adaptively refines its mesh based on a set
of mesh functions.  The maximum cell-to-cell variation of these
functions is maintained around the value of the control
\texttt{mesh\_delta\_coeff}. We use \texttt{mesh\_delta\_coeff}\,=\,1
for all the results reported. Primarily as a result of these two mesh parameters, 
the total number of cells in the baseline
model is $\simeq$\,8,000 cells.

A \MESA\ parameter controlling the time resolution is the 
largest change in the central temperature allowed over a timestep,
\texttt{delta\_lgT\_cntr\_limit}. For all the reported results, we use \texttt{delta\_lgT\_cntr\_limit}\,=\,0.001. 
\MESA\ can also adaptively adjusts the timestep based on other criteria, but this setting
dominates the size of every timestep as the baseline WD model cools.
The total number of timesteps in the baseline model
is $\simeq$\,1,000  and varies roughly linearly with \texttt{delta\_lgT\_cntr\_limit}.

\begin{figure}[!htb]
\centering
\includegraphics[trim={1.0cm 1.5cm 0.0cm 1.2cm},clip,width=1.1\apjcolwidth]{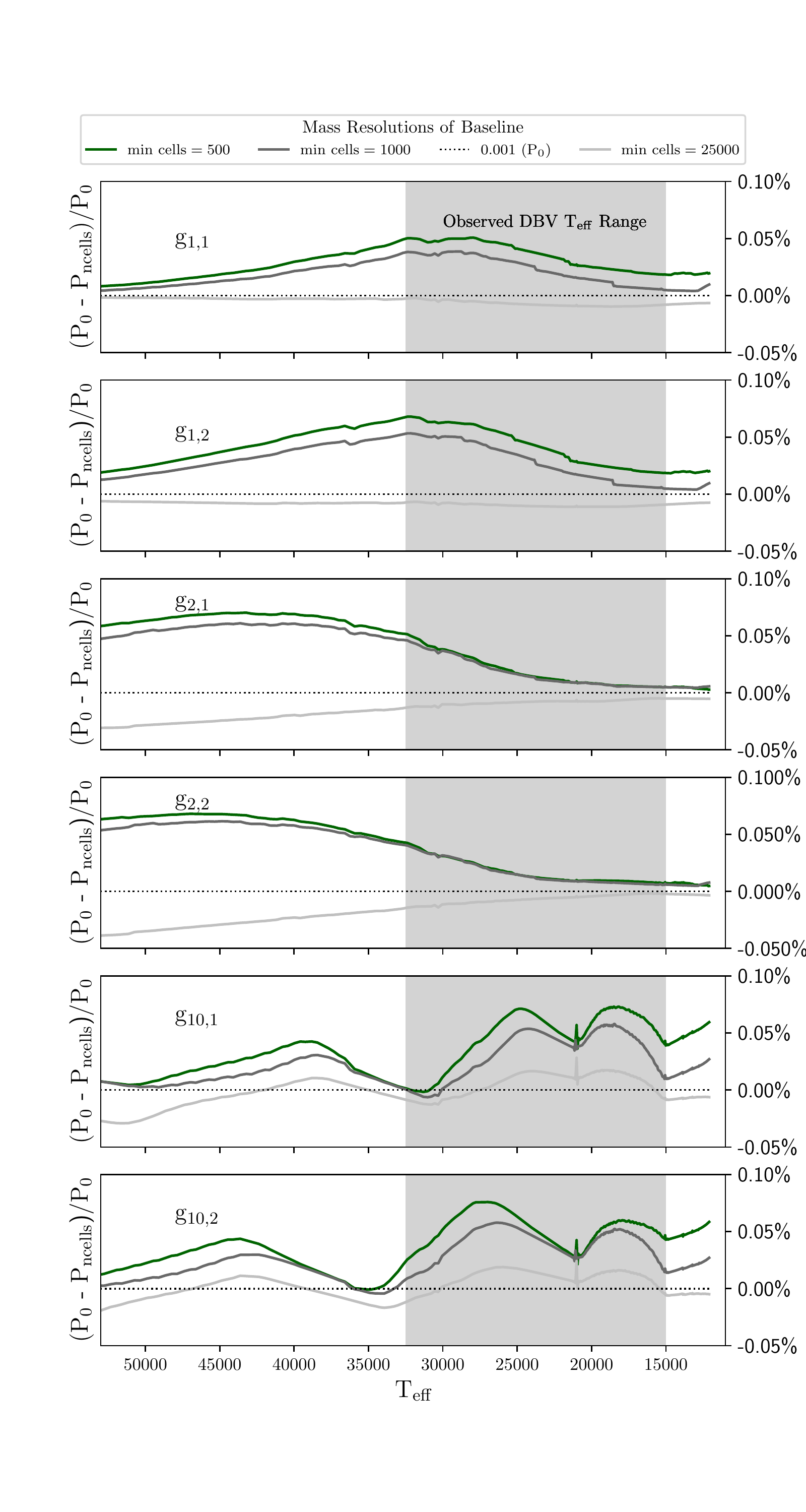}
\caption{Relative differences in the $g_{1,1}$, $g_{1,2}$, $g_{2,1}$, $g_{2,2}$, $g_{10,1}$, $g_{10,2}$ pulsation periods  
for different minimum mass resolutions as the baseline WD models cool.
We use the notation $g_{n,\ell}$ for a g-mode of order $n$ and degree $\ell$.  
The minimum mass resolution of 5,000 cells, used for all the results reported, is shown by the black horizontal lines. 
}
\label{fig:mass_convergence}
\end{figure}

Figure~\ref{fig:mass_convergence} shows changes in the low-order g-mode periods
for different $N_{\rm min \ cells}$ as the models cool.
The time resolution is held fixed at \texttt{delta\_lgT\_cntr\_limit}\,=\,0.001.
Our standard $N_{\rm min \ cells}$\,=\,5,000 baseline model is the basis of the comparison
and shown as the horizontal lines.
A model with 10 times less mass resolution than our standard mass resolution, 
$N_{\rm min \ cells}$\,=\,500, induces maximum relative period changes of 
$\simeq$\,0.05\% at $\simeq$\,30,000~K for $g_{1,1}$,
$\simeq$\,0.07\% at $\simeq$\,35,000~K for $g_{1,2}$,
$\simeq$\,0.07\% at $\simeq$\,45,000~K for $g_{2,1}$, and
$\simeq$\,0.07\% at $\simeq$\,45,000~K for $g_{2,2}$.
A model with 5 times less mass resolution than our standard mass resolution, 
$N_{\rm min \ cells}$\,=\,1,000, reduces these maximum relative period changes by $\simeq$\,20\%.
A model with 5 times more mass resolution than our standard mass resolution, 
$N_{\rm min \ cells}$\,=\,25,000 
causes maximum relative period changes of 0.000022\% at $g_{1,1}$ to 0.028\% at $g_{10,1}$.  
These maximum relative period changes are, respectively, a factor of $\simeq$\,20,000 to 20 smaller 
than the relative period change caused by including or excluding \neon[22].

\begin{figure}[!htb]
\centering
\includegraphics[trim={1.0cm 1.5cm 0.0cm 1.2cm},clip,width=1.1\apjcolwidth]{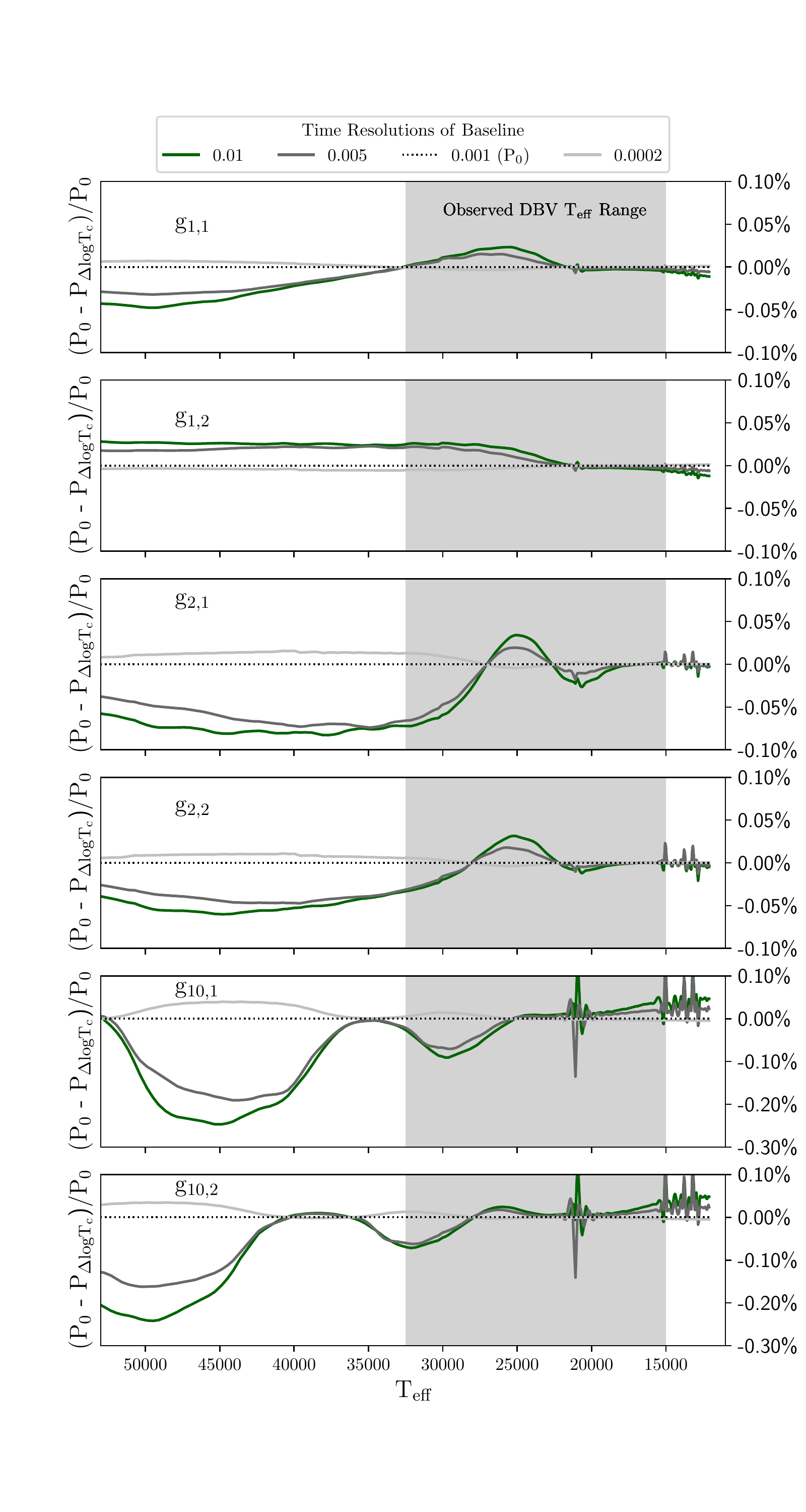}
\caption{Relative differences in the 
$g_{1,1}$, $g_{1,2}$, $g_{2,1}$, $g_{2,2}$, $g_{10,1}$, and $g_{10,2}$
pulsation period for different temporal resolutions as the baseline WD models cool.
The largest change in the central temperature allowed over a timestep,
\texttt{delta\_lgT\_cntr\_limit}\,=\,0.001,
used for all the results reported, is shown by the black horizontal lines.
}
\label{fig:temp_convergence}
\end{figure}

Figure~\ref{fig:temp_convergence} shows changes in the low-order g-mode periods
for different \texttt{delta\_lgT\_cntr\_limit} as the models cool.
The mass resolution is held fixed at $N_{\rm min \ cells}$\,=\,5,000.
Our standard \texttt{delta\_lgT\_cntr\_limit}\,=\,0.001 baseline model is the basis of the comparison
and shown as the horizontal lines.
A model with 10 times less time resolution, \texttt{delta\_lgT\_cntr\_limit}\,=\,0.01, 
causes  maximum relative period changes of 
$\simeq$\,$-$0.05\% at $\simeq$\,50,000~K for $g_{1,1}$,
$\simeq$\,0.02\% at $\simeq$\,50,000~K for $g_{1,2}$,
$\simeq$\,$-$0.06\% at $\simeq$\,40,000~K for $g_{2,1}$,
$\simeq$\,$-$0.05\% at $\simeq$\,45,000~K for $g_{2,2}$, 
$\simeq$\,$-$0.25\% at $\simeq$\,45,000~K for $g_{10,1}$, and
$\simeq$\,$-$0.25\% at $\simeq$\,50,000~K for $g_{10,2}$.
A model with 5 times less time resolution than our standard mass resolution,
\texttt{delta\_lgT\_cntr\_limit}\,=\,0.005,
reduces these maximum relative period changes by $\simeq$\,10\%.
A model with 5 times more time resolution,
\texttt{delta\_lgT\_cntr\_limit}\,=\,0.0002, has average period changes of
0.00061 s for $g_{1,1}$,
$-$0.00077 s for $g_{1,2}$,
0.0034 s for $g_{2,1}$,
0.0010 s for $g_{2,2}$,
0.0021 s for $g_{10,1}$, and 
0.0014 s for $g_{10,2}$.
The average period changes are a factor of $\simeq$\,1000 smaller
than the average period changes caused by including or excluding \neon[22].

\section{Input Physics Details}\label{sec:details}

In this appendix we briefly discuss the salient physics used in our \MESA\ models.

\subsection{Thermodynamics}\label{sec:eos} 

The \MESA\ r12115 equation of state (EOS) is a blend of the OPAL
\citep{rogers_2002_aa}, SCVH \citep{saumon_1995_aa}, PTEH
\citet{pols_1995_aa}, HELM \citep{timmes_2000_ab}, and PC
\citep{potekhin_2010_aa} EOSes. The \MESA\ EOS also covers the
late stages of WD cooling where the ions in the core
crystallize \citep[e.g.,][]{bauer_2020_aa}.  
WD interiors lie in the PC region of the MESA EOS, which provides a semi-analytic EOS treatment for arbitrary composition. The default in MESA version 12115 is to account for each species of ion with mass fraction greater than $10^{-3}$ when calling the PC EOS. Therefore changing the interior composition in a WD model, such as including or excluding $^{22}$Ne, self-consistently changes the thermodynamics.

\subsection{Opacities}\label{sec:kap} 

\MESA\ r12115 divides the radiative opacity tables into two
temperature regimes, low ($T$\,$\lesssim$\,10$^4$ K) and high ($T$\,$\gtrsim$\,10$^4$ K).
For the stellar evolution calculations from the pre-MS to a WD we use the \citet{ferguson_2005_aa} low-temperature regions,
and for the high-temperature regions we use the 
OPAL Type I opacities \citep{iglesias_1996_aa}, smoothly transitioning to the OPAL Type II opacities \citep{iglesias_1996_aa} 
starting at the end of core H-burning.
In our WD models, the radiative opacities are provided by the OPAL Type 2 tables, which are functions of the
hydrogen mass fraction X, metal mass fraction Z, and the C/O-enhancements.  
Thus for the same temperature and density, our $X$(\neon[22]) $\rightarrow$ $X$(\nitrogen[14]) replacement in 
Section~\ref{s.pchanges.n14} does not change the radiative opacities.
Our $X$(\nitrogen[14]) $\rightarrow$ $X$(\helium[4]) and $X$(\neon[22]) with $\rightarrow$ $X$(\carbon[12])
replacements to generate zero-metallicity ab initio DB WD in Section~\ref{s.pchanges.usual}
decreases Z in the He-dominated envelope and increases the C enhancement in the interior.
Conversely, our doubling $X$(\nitrogen[14]) at the expense of $X$(\helium[4])
and doubling $X$(\neon[22]) at the expense of $X$(\carbon[12]) to generate
a super-solar metallicity ab initio DB WD in Section~\ref{s.pchanges.usual}
increases Z in the He-dominated envelope and decreases the C enhancement in the interior.
Electron conduction opacities are from \citet{cassisi_2007_aa},  
which are the relevant opacity in the WD interior. The conduction opacities are a function of 
the mean atomic number $\bar{Z}$, which \MESA\ evaluates using the full composition vector in each cell.

\subsection{Nuclear Reaction Networks}\label{sec:nets}

We use \MESA’s \code{mesa\_49.net}, 
a nuclear reaction network that follows 49 isotopes from 
\hydrogen[1] to \sulfur[34], including \neon[22]. 
This impact of this reaction network on
properties of CO WDs from Monte Carlo stellar models
is discussed by \citet{fields_2016_aa}.
All forward thermonuclear reaction rates are from the JINA reaclib
version V2.2 2017-10-20 \citep{cyburt_2010_ab}. Inverse rates are
calculated directly from the forward rates (those with positive
$Q$-value) using detailed balance, rather than using fitted
rates. The nuclear partition functions used to calculate the inverse rates are
from \citet{rauscher_2000_aa}.  Electron
screening factors for both weak and strong thermonuclear reactions are
from \citet{chugunov_2007_aa} with plasma parameters from
\citet{itoh_1979_aa}.  All the weak rates are based (in order of
precedence) on the tabulations of \citet{langanke_2000_aa},
\citet{oda_1994_aa}, and \citet{fuller_1985_aa}.
Thermal neutrino energy losses are from \citet{itoh_1996_aa}.

\subsection{Mass Loss}\label{sec:mdot}

The implementations of mass loss in \MESA\ r12115 are
based on a number of observationally and theoretically motivated
prescriptions, but uncertainties remain on line-driven and dust-driven winds
\citep{dupree_1986_aa,willson_2000_aa,boulangier_2019_aa}.
We follow the mass loss settings used by the MIST isochrones \citep{dotter_2016_aa, choi_2016_aa},
with a combination of the Reimer mass loss prescription \citep{reimer_1975_aa} with
$\eta$=0.1 on the Red Giant Branch  and a Bl\"ocker mass loss prescription
\citep{blocker_1995_aa} with $\eta$=0.5 on the AGB.

\subsection{Rotation and Magnetic Fields}
\label{sec:rot}

\MESA\ r12115 implements the inherently 3D process of rotation by
making the 1D shellular approximation \citep{zahn_1992_aa,meynet_1997_aa}, 
where the angular velocity is constant over isobars.
The transport of angular momentum and material due to rotationally induced
instabilities is followed using a diffusion approximation
\citep[e.g.,][]{endal_1978_aa, pinsonneault_1989_aa,heger_2000_aa,
maeder_2003_aa,maeder_2004_aa,suijs_2008_aa}
for the dynamical shear instability, 
secular shear instability, 
Eddington-Sweet circulation, 
Goldreich-Schubert-Fricke instability, 
and Spruit-Tayler dynamo.
See \citet{heger_2000_aa} for a description of the 
different instabilities and diffusion coefficients.

Magnetic fields are implemented in \MESA\ using the formalism of
\citet{heger_2005_aa}, where a magnetic torque due to a dynamo
\citep{spruit_2002_aa} allows angular momentum to be transported
inside the star. The azimuthal and radial
components of the magnetic field are modeled as
$B_\phi$\,$\sim$\,$ r \sqrt{(4\pi\rho)}\omega_A$ and $B_r$\,$\sim$\,$
B_\phi/(rk)$ respectively, where $r$ is the radial coordinate, 
$\omega_A$ the Alfv\'en frequency, and $k$ the wavenumber.  These
magnetic fields provide a torque $S$\,=\,$B_rB_\phi/(4\pi)$ which
slows down the rotation rate by decreasing the amount of differential rotation \citep{heger_2005_aa}.

We initialize rotation by
imposing a solid body rotation law, $\rot$\,=\,1.9$\times$10$^{-4}$, at the ZAMS.
ZAMS is defined as where the
nuclear burning luminosity is 99\% of the total luminosity,
and the rotation rate is normalized by the surface critical rotation rate  
$\Omega_{crit}$\,=\,$\sqrt{(1 - L/L_{{\rm edd}}) c M / R^3}$, where $c$ is the speed of light, $M$ is the mass of the star,
$R$ the stellar radius, $L$ the luminosity and $L_{{\rm edd}}$ the Eddington luminosity. 
The initial magnetic field is set to $B_r$\,=\,$B_\phi$\,=\,0. 
Effects from rotationally induced mass loss are not included.

\subsection{Element Diffusion}\label{sec:diffu}

Element diffusion is implemented in \MESA\ r12115 following \citet{thoul_1994_aa},
and described in Section 3 of \citet{paxton_2018_aa}. All isotopes in the
reaction network are categorized into classes according to their
atomic masses, each of which has a representative member whose
properties are used to calculate the diffusion velocities.  Diffusion
coefficients are calculated, by default, according to
\citet{stanton_2016_aa}, whose formalism is based on binary collision
integrals between each pair of species in the plasma.
The diffusion equation is then solved using the total mass fraction within each class.
From the ZAMS to the construction of the DB WD, we use the ten element classes
\hydrogen[1], \helium[3], \helium[4], 
\carbon[12], \nitrogen[14] \oxygen[16], \neon[20], 
\neon[22], \magnesium[24] and \silicon[28].

\clearpage

\bibliographystyle{aasjournal}


\begin{thebibliography}{}
\expandafter\ifx\csname natexlab\endcsname\relax\def\natexlab#1{#1}\fi
\providecommand{\url}[1]{\href{#1}{#1}}
\providecommand{\dodoi}[1]{doi:~\href{http://doi.org/#1}{\nolinkurl{#1}}}
\providecommand{\doeprint}[1]{\href{http://ascl.net/#1}{\nolinkurl{http://ascl.net/#1}}}
\providecommand{\doarXiv}[1]{\href{https://arxiv.org/abs/#1}{\nolinkurl{https://arxiv.org/abs/#1}}}

\bibitem[{{Abbott} {et~al.}(2020){Abbott}, {Abbott}, {Abraham}, {Acernese},
  {Ackley}, {Adams}, {Adhikari}, {Adya}, {Affeldt}, {Agathos}, \&
  et~al.}]{abbott_2020_aa}
{Abbott}, R., {Abbott}, T.~D., {Abraham}, S., {et~al.} 2020, \apjl, 900, L13,
  \dodoi{10.3847/2041-8213/aba493}

\bibitem[{{Alsing} {et~al.}(2020){Alsing}, {Peiris}, {Leja}, {Hahn}, {Tojeiro},
  {Mortlock}, {Leistedt}, {Johnson}, \& {Conroy}}]{alsing_2020_aa}
{Alsing}, J., {Peiris}, H., {Leja}, J., {et~al.} 2020, \apjs, 249, 5,
  \dodoi{10.3847/1538-4365/ab917f}

\bibitem[{{Althaus} \& {Benvenuto}(1997)}]{althaus_1997_aa}
{Althaus}, L.~G., \& {Benvenuto}, O.~G. 1997, \apj, 477, 313,
  \dodoi{10.1086/303686}

\bibitem[{{Althaus} {et~al.}(2020){Althaus}, {Gil Pons}, {C{\'o}rsico}, {Miller
  Bertolami}, {De Ger{\'o}nimo}, {Camisassa}, {Torres}, {Gutierrez}, \&
  {Rebassa-Mansergas}}]{althaus_2020_aa}
{Althaus}, L.~G., {Gil Pons}, P., {C{\'o}rsico}, A.~H., {et~al.} 2020, arXiv
  e-prints, arXiv:2011.10439.
\newblock \doarXiv{2011.10439}

\bibitem[{{Arcones} {et~al.}(2017){Arcones}, {Bardayan}, {Beers}, {Bernstein},
  {Blackmon}, {Messer}, {Brown}, {Brown}, {Brune}, {Champagne}, {Chieffi},
  {Couture}, {Danielewicz}, {Diehl}, {El-Eid}, {Escher}, {Fields},
  {Fr{\"o}hlich}, {Herwig}, {Hix}, {Iliadis}, {Lynch}, {McLaughlin}, {Meyer},
  {Mezzacappa}, {Nunes}, {O'Shea}, {Prakash}, {Pritychenko}, {Reddy}, {Rehm},
  {Rogachev}, {Rutledge}, {Schatz}, {Smith}, {Stairs}, {Steiner}, {Strohmayer},
  {Timmes}, {Townsley}, {Wiescher}, {Zegers}, \& {Zingale}}]{arcones_2017_aa}
{Arcones}, A., {Bardayan}, D.~W., {Beers}, T.~C., {et~al.} 2017, Progress in
  Particle and Nuclear Physics, 94, 1, \dodoi{10.1016/j.ppnp.2016.12.003}

\bibitem[{{Baines} \& {Gill}(1969)}]{baines_1969_aa}
{Baines}, P.~G., \& {Gill}, A.~E. 1969, Journal of Fluid Mechanics, 37, 289,
  \dodoi{10.1017/S0022112069000553}

\bibitem[{{Balona} \& {Ozuyar}(2020)}]{balona_2020_aa}
{Balona}, L.~A., \& {Ozuyar}, D. 2020, \mnras, 493, 5871,
  \dodoi{10.1093/mnras/staa670}

\bibitem[{{Bauer} \& {Bildsten}(2018)}]{bauer_2018_aa}
{Bauer}, E.~B., \& {Bildsten}, L. 2018, \apj, 859, L19,
  \dodoi{10.3847/2041-8213/aac492}

\bibitem[{{Bauer} {et~al.}(2020){Bauer}, {Schwab}, {Bildsten}, \&
  {Cheng}}]{bauer_2020_aa}
{Bauer}, E.~B., {Schwab}, J., {Bildsten}, L., \& {Cheng}, S. 2020, arXiv
  e-prints, arXiv:2009.04025.
\newblock \doarXiv{2009.04025}

\bibitem[{{Becker} \& {Iben}(1979)}]{becker_1979_aa}
{Becker}, S.~A., \& {Iben}, Jr., I. 1979, \apj, 232, 831,
  \dodoi{10.1086/157345}

\bibitem[{{Becker} \& {Iben}(1980)}]{becker_1980_aa}
---. 1980, \apj, 237, 111, \dodoi{10.1086/157850}

\bibitem[{{Bell} {et~al.}(2019){Bell}, {C{\'o}rsico}, {Bischoff-Kim},
  {Althaus}, {Bradley}, {Calcaferro}, {Montgomery}, {Uzundag}, {Baran},
  {Bogn{\'a}r}, {Charpinet}, {Ghasemi}, \& {Hermes}}]{bell_2019}
{Bell}, K.~J., {C{\'o}rsico}, A.~H., {Bischoff-Kim}, A., {et~al.} 2019, \aap,
  632, A42, \dodoi{10.1051/0004-6361/201936340}

\bibitem[{{Bildsten} \& {Hall}(2001)}]{bildsten_2001_aa}
{Bildsten}, L., \& {Hall}, D.~M. 2001, \apjl, 549, L219

\bibitem[{{Bischoff-Kim} \& {Montgomery}(2018)}]{bischoff-kim_2018_aa}
{Bischoff-Kim}, A., \& {Montgomery}, M.~H. 2018, \aj, 155, 187,
  \dodoi{10.3847/1538-3881/aab70e}

\bibitem[{{Bischoff-Kim} {et~al.}(2014){Bischoff-Kim}, {{\O}stensen}, {Hermes},
  \& {Provencal}}]{bischoff-kim_ostensen_2014}
{Bischoff-Kim}, A., {{\O}stensen}, R.~H., {Hermes}, J.~J., \& {Provencal},
  J.~L. 2014, \apj, 794, 39, \dodoi{10.1088/0004-637X/794/1/39}

\bibitem[{{Bischoff-Kim} {et~al.}(2019){Bischoff-Kim}, {Provencal}, {Bradley},
  {Montgomery}, {Shipman}, {Harrold}, {Howard}, {Strickland}, {Chandler},
  {Campbell}, {Arredondo}, {Linn}, {Russell}, {Doyle}, {Brickhouse}, {Peters},
  {Kim}, {Jiang}, {Mao}, {Kusakin}, {Sergeev}, {Andreev}, {Velichko},
  {Janulis}, {Pakstiene}, {Ali{\c{c}}avu{\textcommabelow s}}, {Horoz}, {Zola},
  {Og{\l}oza}, {Koziel-Wierzbowska}, {Kundera}, {Jableka}, {Debski}, {Baran},
  {Meingast}, {Nagel}, {Loebling}, {Heinitz}, {Hoyer}, {Bogn{\'a}r},
  {Castanheira}, \& {Erdem}}]{bischoff-kim_agnes_2019}
{Bischoff-Kim}, A., {Provencal}, J.~L., {Bradley}, P.~A., {et~al.} 2019, \apj,
  871, 13, \dodoi{10.3847/1538-4357/aae2b1}

\bibitem[{{Bl{\"o}cker}(2001)}]{blocker_2001_aa}
{Bl{\"o}cker}, T. 2001, \apss, 275, 1.
\newblock \doarXiv{astro-ph/0102135}

\bibitem[{{Bloecker}(1995{\natexlab{a}})}]{blocker_1995_aa}
{Bloecker}, T. 1995{\natexlab{a}}, \aap, 297, 727

\bibitem[{{Bloecker}(1995{\natexlab{b}})}]{blocker_1995_ab}
---. 1995{\natexlab{b}}, \aap, 299, 755

\bibitem[{{Borexino Collaboration} {et~al.}(2018){Borexino Collaboration},
  {Agostini}, {Altenm{\"u}ller}, {Appel}, {Atroshchenko}, {Bagdasarian},
  {Basilico}, {Bellini}, {Benziger}, {Bick}, {Bonfini}, {Bravo}, {Caccianiga},
  {Calaprice}, {Caminata}, {Caprioli}, {Carlini}, {Cavalcante}, {Chepurnov},
  {Choi}, {Collica}, {D'Angelo}, {Davini}, {Derbin}, {Ding}, {Di Ludovico}, {Di
  Noto}, {Drachnev}, {Fomenko}, {Formozov}, {Franco}, {Gabriele}, {Galbiati},
  {Ghiano}, {Giammarchi}, {Goretti}, {Gromov}, {Guffanti}, {Hagner}, {Houdy},
  {Hungerford}, {Ianni}, {Ianni}, {Jany}, {Jeschke}, {Kobychev}, {Korablev},
  {Korga}, {Kryn}, {Laubenstein}, {Litvinovich}, {Lombardi}, {Lombardi},
  {Ludhova}, {Lukyanchenko}, {Lukyanchenko}, {Machulin}, {Manuzio}, {Marcocci},
  {Martyn}, {Meroni}, {Meyer}, {Miramonti}, {Misiaszek}, {Muratova}, {Neumair},
  {Oberauer}, {Opitz}, {Orekhov}, {Ortica}, {Pallavicini}, {Papp}, {Penek},
  {Pilipenko}, {Pocar}, {Porcelli}, {Raikov}, {Ranucci}, {Razeto}, {Re},
  {Redchuk}, {Romani}, {Roncin}, {Rossi}, {Sch{\"o}nert}, {Semenov},
  {Skorokhvatov}, {Smirnov}, {Sotnikov}, {Stokes}, {Suvorov}, {Tartaglia},
  {Testera}, {Thurn}, {Toropova}, {Unzhakov}, {Villante}, {Vishneva},
  {Vogelaar}, {von Feilitzsch}, {ang}, {Weinz}, {Wojcik}, {Wurm}, {Yokley},
  {Zaimidoroga}, {Zavatarelli}, {Zuber}, \&
  {Zuzel}}]{borexino-collaboration_2018_aa}
{Borexino Collaboration}, {Agostini}, M., {Altenm{\"u}ller}, K., {et~al.} 2018,
  \nat, 562, 505, \dodoi{10.1038/s41586-018-0624-y}

\bibitem[{{Borexino Collaboration} {et~al.}(2020){Borexino Collaboration},
  {Agostini}, {Altenm{\"u}ller}, {Appel}, {Atroshchenko}, {Bagdasarian},
  {Basilico}, {Bellini}, {Benziger}, {Bick}, {Bonfini}, {Bravo}, {Caccianiga},
  {Calaprice}, {Caminata}, {Caprioli}, {Carlini}, {Cavalcante}, {Chepurnov},
  {Choi}, {Collica}, {D'Angelo}, {Davini}, {Derbin}, {Ding}, {Di Ludovico}, {Di
  Noto}, {Drachnev}, {Fomenko}, {Formozov}, {Franco}, {Gabriele}, {Galbiati},
  {Ghiano}, {Giammarchi}, {Goretti}, {Gromov}, {Guffanti}, {Hagner}, {Houdy},
  {Hungerford}, {Ianni}, {Ianni}, {Jany}, {Jeschke}, {Kobychev}, {Korablev},
  {Korga}, {Kryn}, {Laubenstein}, {Litvinovich}, {Lombardi}, {Lombardi},
  {Ludhova}, {Lukyanchenko}, {Lukyanchenko}, {Machulin}, {Manuzio}, {Marcocci},
  {Martyn}, {Meroni}, {Meyer}, {Miramonti}, {Misiaszek}, {Muratova}, {Neumair},
  {Oberauer}, {Opitz}, {Orekhov}, {Ortica}, {Pallavicini}, {Papp}, {Penek},
  {Pilipenko}, {Pocar}, {Porcelli}, {Raikov}, {Ranucci}, {Razeto}, {Re},
  {Redchuk}, {Romani}, {Roncin}, {Rossi}, {Sch{\"o}nert}, {Semenov},
  {Skorokhvatov}, {Smirnov}, {Sotnikov}, {Stokes}, {Suvorov}, {Tartaglia},
  {Testera}, {Thurn}, {Toropova}, {Unzhakov}, {Villante}, {Vishneva},
  {Vogelaar}, {von Feilitzsch}, {ang}, {Weinz}, {Wojcik}, {Wurm}, {Yokley},
  {Zaimidoroga}, {Zavatarelli}, {Zuber}, \&
  {Zuzel}}]{borexino-collaboration_2020_aa}
---. 2020, \nat, 587, 577

\bibitem[{{Boulangier} {et~al.}(2019){Boulangier}, {Clementel}, {van Marle},
  {Decin}, \& {de Koter}}]{boulangier_2019_aa}
{Boulangier}, J., {Clementel}, N., {van Marle}, A.~J., {Decin}, L., \& {de
  Koter}, A. 2019, \mnras, 482, 5052, \dodoi{10.1093/mnras/sty2560}

\bibitem[{{Bravo} {et~al.}(1992){Bravo}, {Isern}, {Canal}, \&
  {Labay}}]{bravo_1992_aa}
{Bravo}, E., {Isern}, J., {Canal}, R., \& {Labay}, J. 1992, \aap, 257, 534

\bibitem[{{Brown} {et~al.}(2013){Brown}, {Garaud}, \&
  {Stellmach}}]{brown_2013_ab}
{Brown}, J.~M., {Garaud}, P., \& {Stellmach}, S. 2013, \apj, 768, 34,
  \dodoi{10.1088/0004-637X/768/1/34}

\bibitem[{{Buchler} \& {Yueh}(1976)}]{buchler_1976_aa}
{Buchler}, J.~R., \& {Yueh}, W.~R. 1976, \apj, 210, 440, \dodoi{10.1086/154847}

\bibitem[{{Camisassa} {et~al.}(2016){Camisassa}, {Althaus}, {C{\'o}rsico},
  {Vinyoles}, {Serenelli}, {Isern}, {Miller Bertolami}, \&
  {Garc{\'\i}a─Berro}}]{camisassa_2016_aa}
{Camisassa}, M.~E., {Althaus}, L.~G., {C{\'o}rsico}, A.~H., {et~al.} 2016,
  \apj, 823, 158, \dodoi{10.3847/0004-637X/823/2/158}

\bibitem[{{Cassisi} {et~al.}(2007){Cassisi}, {Potekhin}, {Pietrinferni},
  {Catelan}, \& {Salaris}}]{cassisi_2007_aa}
{Cassisi}, S., {Potekhin}, A.~Y., {Pietrinferni}, A., {Catelan}, M., \&
  {Salaris}, M. 2007, \apj, 661, 1094

\bibitem[{{Charpinet} {et~al.}(2019){Charpinet}, {Brassard}, {Giammichele}, \&
  {Fontaine}}]{charpinet_2019_aa}
{Charpinet}, S., {Brassard}, P., {Giammichele}, N., \& {Fontaine}, G. 2019,
  \aap, 628, L2, \dodoi{10.1051/0004-6361/201935823}

\bibitem[{{Cheng} {et~al.}(2019){Cheng}, {Cummings}, \&
  {M{\'e}nard}}]{cheng_2019_aa}
{Cheng}, S., {Cummings}, J.~D., \& {M{\'e}nard}, B. 2019, \apj, 886, 100,
  \dodoi{10.3847/1538-4357/ab4989}

\bibitem[{{Choi} {et~al.}(2016){Choi}, {Dotter}, {Conroy}, {Cantiello},
  {Paxton}, \& {Johnson}}]{choi_2016_aa}
{Choi}, J., {Dotter}, A., {Conroy}, C., {et~al.} 2016, \apj, 823, 102,
  \dodoi{10.3847/0004-637X/823/2/102}

\bibitem[{{Chugunov} {et~al.}(2007){Chugunov}, {Dewitt}, \&
  {Yakovlev}}]{chugunov_2007_aa}
{Chugunov}, A.~I., {Dewitt}, H.~E., \& {Yakovlev}, D.~G. 2007, \prd, 76,
  025028, \dodoi{10.1103/PhysRevD.76.025028}

\bibitem[{{Constantino} {et~al.}(2015){Constantino}, {Campbell},
  {Christensen-Dalsgaard}, {Lattanzio}, \& {Stello}}]{constantino_2015_aa}
{Constantino}, T., {Campbell}, S.~W., {Christensen-Dalsgaard}, J., {Lattanzio},
  J.~C., \& {Stello}, D. 2015, \mnras, 452, 123, \dodoi{10.1093/mnras/stv1264}

\bibitem[{{Constantino} {et~al.}(2017){Constantino}, {Campbell}, \&
  {Lattanzio}}]{constantino_2017_aa}
{Constantino}, T., {Campbell}, S.~W., \& {Lattanzio}, J.~C. 2017, \mnras, 472,
  4900, \dodoi{10.1093/mnras/stx2321}

\bibitem[{{Constantino} {et~al.}(2016){Constantino}, {Campbell}, {Lattanzio},
  \& {van Duijneveldt}}]{constantino_2016_aa}
{Constantino}, T., {Campbell}, S.~W., {Lattanzio}, J.~C., \& {van Duijneveldt},
  A. 2016, \mnras, 456, 3866, \dodoi{10.1093/mnras/stv2939}

\bibitem[{{C{\'o}rsico} {et~al.}(2019){C{\'o}rsico}, {Althaus}, {Miller
  Bertolami}, \& {Kepler}}]{corsico_2019_aa}
{C{\'o}rsico}, A.~H., {Althaus}, L.~G., {Miller Bertolami}, M.~M., \& {Kepler},
  S.~O. 2019, \aapr, 27, 7, \dodoi{10.1007/s00159-019-0118-4}

\bibitem[{{Cyburt} {et~al.}(2010){Cyburt}, {Amthor}, {Ferguson}, {Meisel},
  {Smith}, {Warren}, {Heger}, {Hoffman}, {Rauscher}, {Sakharuk}, {Schatz},
  {Thielemann}, \& {Wiescher}}]{cyburt_2010_ab}
{Cyburt}, R.~H., {Amthor}, A.~M., {Ferguson}, R., {et~al.} 2010, \apjs, 189,
  240, \dodoi{10.1088/0067-0049/189/1/240}

\bibitem[{{D'Antona} \& {Mazzitelli}(1990)}]{dantona_1990_aa}
{D'Antona}, F., \& {Mazzitelli}, I. 1990, \araa, 28, 139,
  \dodoi{10.1146/annurev.aa.28.090190.001035}

\bibitem[{{De Ger{\'o}nimo} {et~al.}(2019){De Ger{\'o}nimo}, {Battich}, {Miller
  Bertolami}, {Althaus}, \& {C{\'o}rsico}}]{de-geronimo_2019_aa}
{De Ger{\'o}nimo}, F.~C., {Battich}, T., {Miller Bertolami}, M.~M., {Althaus},
  L.~G., \& {C{\'o}rsico}, A.~H. 2019, \aap, 630, A100,
  \dodoi{10.1051/0004-6361/201834988}

\bibitem[{{deBoer} {et~al.}(2017){deBoer}, {G{\"o}rres}, {Wiescher}, {Azuma},
  {Best}, {Brune}, {Fields}, {Jones}, {Pignatari}, {Sayre}, {Smith}, {Timmes},
  \& {Uberseder}}]{deboer_2017_aa}
{deBoer}, R.~J., {G{\"o}rres}, J., {Wiescher}, M., {et~al.} 2017, Reviews of
  Modern Physics, 89, 035007, \dodoi{10.1103/RevModPhys.89.035007}

\bibitem[{{Deloye} \& {Bildsten}(2002)}]{deloye_2002_aa}
{Deloye}, C.~J., \& {Bildsten}, L. 2002, \apj, 580, 1077

\bibitem[{{Demarque} \& {Mengel}(1971)}]{demarque_1971_aa}
{Demarque}, P., \& {Mengel}, J.~G. 1971, \apj, 164, 317, \dodoi{10.1086/150841}

\bibitem[{{Denissenkov} {et~al.}(2013){Denissenkov}, {Herwig}, {Truran}, \&
  {Paxton}}]{denissenkov_2013_aa}
{Denissenkov}, P.~A., {Herwig}, F., {Truran}, J.~W., \& {Paxton}, B. 2013,
  \apj, 772, 37, \dodoi{10.1088/0004-637X/772/1/37}

\bibitem[{{Dotter}(2016)}]{dotter_2016_aa}
{Dotter}, A. 2016, \apjs, 222, 8, \dodoi{10.3847/0067-0049/222/1/8}

\bibitem[{{Dufour} {et~al.}(2017){Dufour}, {Blouin}, {Coutu},
  {Fortin-Archambault}, {Thibeault}, {Bergeron}, \&
  {Fontaine}}]{dufour_2017_aa}
{Dufour}, P., {Blouin}, S., {Coutu}, S., {et~al.} 2017, in Astronomical Society
  of the Pacific Conference Series, Vol. 509, 20th European White Dwarf
  Workshop, ed. P.~E. {Tremblay}, B.~{Gaensicke}, \& T.~{Marsh}, 3.
\newblock \doarXiv{1610.00986}

\bibitem[{{Dupree}(1986)}]{dupree_1986_aa}
{Dupree}, A.~K. 1986, \araa, 24, 377,
  \dodoi{10.1146/annurev.aa.24.090186.002113}

\bibitem[{{Dziembowski}(1971)}]{dziembowski_1971_aa}
{Dziembowski}, W.~A. 1971, \actaa, 21, 289

\bibitem[{{Endal} \& {Sofia}(1978)}]{endal_1978_aa}
{Endal}, A.~S., \& {Sofia}, S. 1978, \apj, 220, 279

\bibitem[{{Farag} {et~al.}(2020){Farag}, {Timmes}, {Taylor}, {Patton}, \&
  {Farmer}}]{farag_2020_aa}
{Farag}, E., {Timmes}, F.~X., {Taylor}, M., {Patton}, K.~M., \& {Farmer}, R.
  2020, \apj, 893, 133, \dodoi{10.3847/1538-4357/ab7f2c}

\bibitem[{{Farmer} {et~al.}(2015){Farmer}, {Fields}, \&
  {Timmes}}]{farmer_2015_aa}
{Farmer}, R., {Fields}, C.~E., \& {Timmes}, F.~X. 2015, \apj, 807, 184,
  \dodoi{10.1088/0004-637X/807/2/184}

\bibitem[{{Farmer} {et~al.}(2020){Farmer}, {Renzo}, {de Mink}, {Fishbach}, \&
  {Justham}}]{farmer_2020_aa}
{Farmer}, R., {Renzo}, M., {de Mink}, S., {Fishbach}, M., \& {Justham}, S.
  2020, arXiv e-prints, arXiv:2006.06678.
\newblock \doarXiv{2006.06678}

\bibitem[{{Ferguson} {et~al.}(2005){Ferguson}, {Alexander}, {Allard}, {Barman},
  {Bodnarik}, {Hauschildt}, {Heffner-Wong}, \& {Tamanai}}]{ferguson_2005_aa}
{Ferguson}, J.~W., {Alexander}, D.~R., {Allard}, F., {et~al.} 2005, \apj, 623,
  585, \dodoi{10.1086/428642}

\bibitem[{{Fields} {et~al.}(2016){Fields}, {Farmer}, {Petermann}, {Iliadis}, \&
  {Timmes}}]{fields_2016_aa}
{Fields}, C.~E., {Farmer}, R., {Petermann}, I., {Iliadis}, C., \& {Timmes},
  F.~X. 2016, \apj, 823, 46, \dodoi{10.3847/0004-637X/823/1/46}

\bibitem[{{Fontaine} \& {Brassard}(2002)}]{fontaine_2002_aa}
{Fontaine}, G., \& {Brassard}, P. 2002, \apjl, 581, L33, \dodoi{10.1086/345787}

\bibitem[{{Fontaine} \& {Brassard}(2008)}]{fontaine_2008_aa}
---. 2008, \pasp, 120, 1043, \dodoi{10.1086/592788}

\bibitem[{{Freedman} {et~al.}(2008){Freedman}, {Marley}, \&
  {Lodders}}]{freedman_2008_aa}
{Freedman}, R.~S., {Marley}, M.~S., \& {Lodders}, K. 2008, \apjs, 174, 504,
  \dodoi{10.1086/521793}

\bibitem[{{Frommhold} {et~al.}(2010){Frommhold}, {Abel}, {Wang}, {Gustafsson},
  {Li}, \& {Hunt}}]{frommhold_2010_aa}
{Frommhold}, L., {Abel}, M., {Wang}, F., {et~al.} 2010, Molecular Physics, 108,
  2265, \dodoi{10.1080/00268976.2010.507556}

\bibitem[{{Fuller} {et~al.}(1985){Fuller}, {Fowler}, \&
  {Newman}}]{fuller_1985_aa}
{Fuller}, G.~M., {Fowler}, W.~A., \& {Newman}, M.~J. 1985, \apj, 293, 1,
  \dodoi{10.1086/163208}

\bibitem[{{Garaud}(2018)}]{garaud_2018_aa}
{Garaud}, P. 2018, Annual Review of Fluid Mechanics, 50, 275,
  \dodoi{10.1146/annurev-fluid-122316-045234}

\bibitem[{{Garc{\'\i}a-Berro} {et~al.}(1997){Garc{\'\i}a-Berro}, {Ritossa}, \&
  {Iben}}]{garcia-berro_1997_aa}
{Garc{\'\i}a-Berro}, E., {Ritossa}, C., \& {Iben}, Jr., I. 1997, \apj, 485,
  765.
\newblock \url{http://stacks.iop.org/0004-637X/485/i=2/a=765}

\bibitem[{{Gautschy}(2012)}]{gautschy_2012_aa}
{Gautschy}, A. 2012, ArXiv e-prints.
\newblock \doarXiv{1208.3870}

\bibitem[{{Giacobbo} \& {Mapelli}(2018)}]{giacobbo_2018_aa}
{Giacobbo}, N., \& {Mapelli}, M. 2018, \mnras, 480, 2011,
  \dodoi{10.1093/mnras/sty1999}

\bibitem[{{Giammichele} {et~al.}(2017){Giammichele}, {Charpinet}, {Brassard},
  \& {Fontaine}}]{giammichele_2017_aa}
{Giammichele}, N., {Charpinet}, S., {Brassard}, P., \& {Fontaine}, G. 2017,
  \aap, 598, A109, \dodoi{10.1051/0004-6361/201629935}

\bibitem[{{Giammichele} {et~al.}(2018){Giammichele}, {Charpinet}, {Fontaine},
  {Brassard}, {Green}, {Van Grootel}, {Bergeron}, {Zong}, \&
  {Dupret}}]{giammichele_2018_aa}
{Giammichele}, N., {Charpinet}, S., {Fontaine}, G., {et~al.} 2018, \nat, 554,
  73, \dodoi{10.1038/nature25136}

\bibitem[{{Hansen} \& {Kawaler}(1994)}]{hansen_1994sipp.book.....H}
{Hansen}, C.~J., \& {Kawaler}, S.~D. 1994, {Stellar Interiors. Physical
  Principles, Structure, and Evolution.} (New York: Springer-Verlag),
  \dodoi{10.1007/978-1-4419-9110-2}

\bibitem[{{Heger} {et~al.}(2000){Heger}, {Langer}, \&
  {Woosley}}]{heger_2000_aa}
{Heger}, A., {Langer}, N., \& {Woosley}, S.~E. 2000, \apj, 528, 368

\bibitem[{{Heger} {et~al.}(2005){Heger}, {Woosley}, \&
  {Spruit}}]{heger_2005_aa}
{Heger}, A., {Woosley}, S.~E., \& {Spruit}, H.~C. 2005, \apj, 626, 350

\bibitem[{{Hekker} \& {Christensen-Dalsgaard}(2017)}]{hekker_2017_aa}
{Hekker}, S., \& {Christensen-Dalsgaard}, J. 2017, \aapr, 25, 1,
  \dodoi{10.1007/s00159-017-0101-x}

\bibitem[{{Hermes} {et~al.}(2017){Hermes}, {Kawaler}, {Bischoff-Kim},
  {Provencal}, {Dunlap}, \& {Clemens}}]{hermes_2017_ab}
{Hermes}, J.~J., {Kawaler}, S.~D., {Bischoff-Kim}, A., {et~al.} 2017, \apj,
  835, 277, \dodoi{10.3847/1538-4357/835/2/277}

\bibitem[{{Herwig}(2005)}]{herwig_2005_aa}
{Herwig}, F. 2005, \araa, 43, 435

\bibitem[{{Hon} {et~al.}(2018){Hon}, {Stello}, \& {Zinn}}]{hon_2018_aa}
{Hon}, M., {Stello}, D., \& {Zinn}, J.~C. 2018, \apj, 859, 64,
  \dodoi{10.3847/1538-4357/aabfdb}

\bibitem[{Hunter(2007)}]{hunter_2007_aa}
Hunter, J.~D. 2007, Computing In Science \&amp; Engineering, 9, 90

\bibitem[{{Iben} \& {Renzini}(1983)}]{iben_1983_aa}
{Iben}, Jr., I., \& {Renzini}, A. 1983, \araa, 21, 271,
  \dodoi{10.1146/annurev.aa.21.090183.001415}

\bibitem[{{Iglesias} \& {Rogers}(1996)}]{iglesias_1996_aa}
{Iglesias}, C.~A., \& {Rogers}, F.~J. 1996, \apj, 464, 943

\bibitem[{{Isern} {et~al.}(1991){Isern}, {Hernanz}, {Mochkovitch}, \&
  {Garcia-Berro}}]{isern_1991_aa}
{Isern}, J., {Hernanz}, M., {Mochkovitch}, R., \& {Garcia-Berro}, E. 1991,
  \aap, 241, L29

\bibitem[{{Itoh} {et~al.}(1996){Itoh}, {Hayashi}, {Nishikawa}, \&
  {Kohyama}}]{itoh_1996_aa}
{Itoh}, N., {Hayashi}, H., {Nishikawa}, A., \& {Kohyama}, Y. 1996, \apjs, 102,
  411

\bibitem[{{Itoh} {et~al.}(1979){Itoh}, {Totsuji}, {Ichimaru}, \&
  {Dewitt}}]{itoh_1979_aa}
{Itoh}, N., {Totsuji}, H., {Ichimaru}, S., \& {Dewitt}, H.~E. 1979, \apj, 234,
  1079

\bibitem[{{Jones} {et~al.}(2013){Jones}, {Hirschi}, {Nomoto}, {Fischer},
  {Timmes}, {Herwig}, {Paxton}, {Toki}, {Suzuki}, {Mart{\'{\i}}nez-Pinedo},
  {Lam}, \& {Bertolli}}]{jones_2013_aa}
{Jones}, S., {Hirschi}, R., {Nomoto}, K., {et~al.} 2013, \apj, 772, 150,
  \dodoi{10.1088/0004-637X/772/2/150}

\bibitem[{{Karakas} \& {Lattanzio}(2014)}]{karakas_2014_aa}
{Karakas}, A.~I., \& {Lattanzio}, J.~C. 2014, ArXiv e-prints.
\newblock \doarXiv{1405.0062}

\bibitem[{{Kawaler} {et~al.}(1985){Kawaler}, {Winget}, \&
  {Hansen}}]{kawaler_1985_aa}
{Kawaler}, S.~D., {Winget}, D.~E., \& {Hansen}, C.~J. 1985, \apj, 295, 547,
  \dodoi{10.1086/163398}

\bibitem[{{Koester}(2010)}]{koester_2010_aa}
{Koester}, D. 2010, \memsai, 81, 921

\bibitem[{{Kutter} \& {Savedoff}(1969)}]{kutter_1969_aa}
{Kutter}, G.~S., \& {Savedoff}, M.~P. 1969, \apj, 156, 1021,
  \dodoi{10.1086/150033}

\bibitem[{{Langanke} \& {Mart{\'{\i}}nez-Pinedo}(2000)}]{langanke_2000_aa}
{Langanke}, K., \& {Mart{\'{\i}}nez-Pinedo}, G. 2000, Nuclear Physics A, 673,
  481, \dodoi{10.1016/S0375-9474(00)00131-7}

\bibitem[{{Lecoanet} {et~al.}(2016){Lecoanet}, {Schwab}, {Quataert},
  {Bildsten}, {Timmes}, {Burns}, {Vasil}, {Oishi}, \&
  {Brown}}]{lecoanet_2016_aa}
{Lecoanet}, D., {Schwab}, J., {Quataert}, E., {et~al.} 2016, \apj, 832, 71,
  \dodoi{10.3847/0004-637X/832/1/71}

\bibitem[{{Lederer} \& {Aringer}(2009)}]{lederer_2009_aa}
{Lederer}, M.~T., \& {Aringer}, B. 2009, \aap, 494, 403,
  \dodoi{10.1051/0004-6361:200810576}

\bibitem[{{Maeder} \& {Meynet}(2003)}]{maeder_2003_aa}
{Maeder}, A., \& {Meynet}, G. 2003, \aap, 411, 543

\bibitem[{{Maeder} \& {Meynet}(2004)}]{maeder_2004_aa}
---. 2004, \aap, 422, 225

\bibitem[{{Marchant} \& {Moriya}(2020)}]{marchant_2020_aa}
{Marchant}, P., \& {Moriya}, T.~J. 2020, \aap, 640, L18,
  \dodoi{10.1051/0004-6361/202038902}

\bibitem[{{Marigo} \& {Aringer}(2009)}]{marigo_2009_aa}
{Marigo}, P., \& {Aringer}, B. 2009, \aap, 508, 1539,
  \dodoi{10.1051/0004-6361/200912598}

\bibitem[{{Metcalfe}(2003)}]{metcalfe_2003_aa}
{Metcalfe}, T.~S. 2003, \apjl, 587, L43, \dodoi{10.1086/375044}

\bibitem[{{Metcalfe} {et~al.}(2003){Metcalfe}, {Montgomery}, \&
  {Kawaler}}]{metcalfe_2003_ab}
{Metcalfe}, T.~S., {Montgomery}, M.~H., \& {Kawaler}, S.~D. 2003, \mnras, 344,
  L88, \dodoi{10.1046/j.1365-8711.2003.07128.x}

\bibitem[{{Metcalfe} {et~al.}(2002){Metcalfe}, {Salaris}, \&
  {Winget}}]{metcalfe_2002_aa}
{Metcalfe}, T.~S., {Salaris}, M., \& {Winget}, D.~E. 2002, \apj, 573, 803,
  \dodoi{10.1086/340796}

\bibitem[{{Meynet} \& {Maeder}(1997)}]{meynet_1997_aa}
{Meynet}, G., \& {Maeder}, A. 1997, \aap, 321, 465

\bibitem[{{Miles} {et~al.}(2016){Miles}, {van Rossum}, {Townsley}, {Timmes},
  {Jackson}, {Calder}, \& {Brown}}]{miles_2016_aa}
{Miles}, B.~J., {van Rossum}, D.~R., {Townsley}, D.~M., {et~al.} 2016, \apj,
  824, 59, \dodoi{10.3847/0004-637X/824/1/59}

\bibitem[{{Mukhopadhyay} {et~al.}(2020){Mukhopadhyay}, {Lunardini}, {Timmes},
  \& {Zuber}}]{mukhopadhyay_2020_aa}
{Mukhopadhyay}, M., {Lunardini}, C., {Timmes}, F.~X., \& {Zuber}, K. 2020,
  \apj, 899, 153, \dodoi{10.3847/1538-4357/ab99a6}

\bibitem[{{Oda} {et~al.}(1994){Oda}, {Hino}, {Muto}, {Takahara}, \&
  {Sato}}]{oda_1994_aa}
{Oda}, T., {Hino}, M., {Muto}, K., {Takahara}, M., \& {Sato}, K. 1994, Atomic
  Data and Nuclear Data Tables, 56, 231, \dodoi{10.1006/adnd.1994.1007}

\bibitem[{{Parsons} {et~al.}(2016){Parsons}, {Rebassa-Mansergas}, {Schreiber},
  {G{\"a}nsicke}, {Zorotovic}, \& {Ren}}]{parsons_2016_aa}
{Parsons}, S.~G., {Rebassa-Mansergas}, A., {Schreiber}, M.~R., {et~al.} 2016,
  \mnras, 463, 2125, \dodoi{10.1093/mnras/stw2143}

\bibitem[{{Patton} {et~al.}(2017){Patton}, {Lunardini}, {Farmer}, \&
  {Timmes}}]{patton_2017_ab}
{Patton}, K.~M., {Lunardini}, C., {Farmer}, R.~J., \& {Timmes}, F.~X. 2017,
  \apj, 851, 6, \dodoi{10.3847/1538-4357/aa95c4}

\bibitem[{{Paxton} {et~al.}(2011){Paxton}, {Bildsten}, {Dotter}, {Herwig},
  {Lesaffre}, \& {Timmes}}]{paxton_2011_aa}
{Paxton}, B., {Bildsten}, L., {Dotter}, A., {et~al.} 2011, \apjs, 192, 3,
  \dodoi{10.1088/0067-0049/192/1/3}

\bibitem[{{Paxton} {et~al.}(2013){Paxton}, {Cantiello}, {Arras}, {Bildsten},
  {Brown}, {Dotter}, {Mankovich}, {Montgomery}, {Stello}, {Timmes}, \&
  {Townsend}}]{paxton_2013_aa}
{Paxton}, B., {Cantiello}, M., {Arras}, P., {et~al.} 2013, \apjs, 208

\bibitem[{{Paxton} {et~al.}(2015){Paxton}, {Marchant}, {Schwab}, {Bauer},
  {Bildsten}, {Cantiello}, {Dessart}, {Farmer}, {Hu}, {Langer}, {Townsend},
  {Townsley}, \& {Timmes}}]{paxton_2015_aa}
{Paxton}, B., {Marchant}, P., {Schwab}, J., {et~al.} 2015, \apjs, 220, 15,
  \dodoi{10.1088/0067-0049/220/1/15}

\bibitem[{{Paxton} {et~al.}(2018){Paxton}, {Schwab}, {Bauer}, {Bildsten},
  {Blinnikov}, {Duffell}, {Farmer}, {Goldberg}, {Marchant}, {Sorokina},
  {Thoul}, {Townsend}, \& {Timmes}}]{paxton_2018_aa}
{Paxton}, B., {Schwab}, J., {Bauer}, E.~B., {et~al.} 2018, \apjs, 234, 34,
  \dodoi{10.3847/1538-4365/aaa5a8}

\bibitem[{{Paxton} {et~al.}(2019){Paxton}, {Smolec}, {Schwab}, {Gautschy},
  {Bildsten}, {Cantiello}, {Dotter}, {Farmer}, {Goldberg}, {Jermyn}, {Kanbur},
  {Marchant}, {Thoul}, {Townsend}, {Wolf}, {Zhang}, \&
  {Timmes}}]{paxton_2019_aa}
{Paxton}, B., {Smolec}, R., {Schwab}, J., {et~al.} 2019, \apjs, 243, 10,
  \dodoi{10.3847/1538-4365/ab2241}

\bibitem[{{Pedersen} {et~al.}(2019){Pedersen}, {Chowdhury}, {Johnston},
  {Bowman}, {Aerts}, {Handler}, {De Cat}, {Neiner}, {David-Uraz}, {Buzasi},
  {Tkachenko}, {Sim{\'o}n-D{\'\i}az}, {Moravveji}, {Sikora}, {Mirouh},
  {Lovekin}, {Cantiello}, {Daszy{\'n}ska-Daszkiewicz}, {Pigulski},
  {Vanderspek}, \& {Ricker}}]{pedersen_2019_aa}
{Pedersen}, M.~G., {Chowdhury}, S., {Johnston}, C., {et~al.} 2019, \apjl, 872,
  L9, \dodoi{10.3847/2041-8213/ab01e1}

\bibitem[{{Pinsonneault} {et~al.}(1989){Pinsonneault}, {Kawaler}, {Sofia}, \&
  {Demarque}}]{pinsonneault_1989_aa}
{Pinsonneault}, M.~H., {Kawaler}, S.~D., {Sofia}, S., \& {Demarque}, P. 1989,
  \apj, 338, 424

\bibitem[{{Placco} {et~al.}(2020){Placco}, {Santucci}, {Yuan}, {Mardini},
  {Holmbeck}, {Wang}, {Surman}, {Hansen}, {Roederer}, {Beers}, {Choplin}, {Ji},
  {Ezzeddine}, {Frebel}, {Sakari}, {Whitten}, \& {Zepeda}}]{placco_2020_aa}
{Placco}, V.~M., {Santucci}, R.~M., {Yuan}, Z., {et~al.} 2020, arXiv e-prints,
  arXiv:2006.04538.
\newblock \doarXiv{2006.04538}

\bibitem[{{Pols} {et~al.}(1995){Pols}, {Tout}, {Eggleton}, \&
  {Han}}]{pols_1995_aa}
{Pols}, O.~R., {Tout}, C.~A., {Eggleton}, P.~P., \& {Han}, Z. 1995, \mnras,
  274, 964, \dodoi{10.1093/mnras/274.3.964}

\bibitem[{{Potekhin} \& {Chabrier}(2010)}]{potekhin_2010_aa}
{Potekhin}, A.~Y., \& {Chabrier}, G. 2010, Contributions to Plasma Physics, 50,
  82, \dodoi{10.1002/ctpp.201010017}

\bibitem[{{Prada Moroni} \& {Straniero}(2009)}]{prada-moroni_2009_aa}
{Prada Moroni}, P.~G., \& {Straniero}, O. 2009, \aap, 507, 1575,
  \dodoi{10.1051/0004-6361/200912847}

\bibitem[{{Rauscher} \& {Thielemann}(2000)}]{rauscher_2000_aa}
{Rauscher}, T., \& {Thielemann}, F.-K. 2000, Atomic Data and Nuclear Data
  Tables, 75, 1, \dodoi{10.1006/adnd.2000.0834}

\bibitem[{{Reimers}(1975)}]{reimer_1975_aa}
{Reimers}, D. 1975, Memoires of the Societe Royale des Sciences de Liege, 8,
  369

\bibitem[{{Rogers} \& {Nayfonov}(2002)}]{rogers_2002_aa}
{Rogers}, F.~J., \& {Nayfonov}, A. 2002, \apj, 576, 1064

\bibitem[{{Rose} {et~al.}(2020){Rose}, {Rubin}, {Cikota}, {Deustua}, {Dixon},
  {Fruchter}, {Jones}, {Riess}, \& {Scolnic}}]{rose_2020_aa}
{Rose}, B.~M., {Rubin}, D., {Cikota}, A., {et~al.} 2020, \apjl, 896, L4,
  \dodoi{10.3847/2041-8213/ab94ad}

\bibitem[{{Saumon} {et~al.}(1995){Saumon}, {Chabrier}, \& {van
  Horn}}]{saumon_1995_aa}
{Saumon}, D., {Chabrier}, G., \& {van Horn}, H.~M. 1995, \apjs, 99, 713,
  \dodoi{10.1086/192204}

\bibitem[{{Seaton}(2005)}]{seaton_2005_aa}
{Seaton}, M.~J. 2005, \mnras, 362, L1, \dodoi{10.1111/j.1365-2966.2005.00019.x}

\bibitem[{{Serenelli} \& {Fukugita}(2005)}]{serenelli_2005_aa}
{Serenelli}, A.~M., \& {Fukugita}, M. 2005, \apjl, 632, L33,
  \dodoi{10.1086/497535}

\bibitem[{{Sim{\'o}n-D{\'\i}az} {et~al.}(2018){Sim{\'o}n-D{\'\i}az}, {Aerts},
  {Urbaneja}, {Camacho}, {Antoci}, {Fredslund Andersen}, {Grundahl}, \&
  {Pall{\'e}}}]{simon-diaz_2018_aa}
{Sim{\'o}n-D{\'\i}az}, S., {Aerts}, C., {Urbaneja}, M.~A., {et~al.} 2018, \aap,
  612, A40, \dodoi{10.1051/0004-6361/201732160}

\bibitem[{{Simpson} {et~al.}(2019){Simpson}, {Abe}, {Bronner}, {Hayato},
  {Ikeda}, {Ito}, {Iyogi}, {Kameda}, {Kataoka}, {Kato}, {Kishimoto}, {Marti},
  {Miura}, {Moriyama}, {Mochizuki}, {Nakahata}, {Nakajima}, {Nakayama},
  {Okada}, {Okamoto}, {Orii}, {Pronost}, {Sekiya}, {Shiozawa}, {Sonoda},
  {Takeda}, {Takenaka}, {Tanaka}, {Yano}, {Akutsu}, {Kajita}, {Okumura},
  {Wang}, {Xia}, {Bravo-Bergu{\~n}o}, {Labarga}, {Fernandez}, {Blaszczyk},
  {Kachulis}, {Kearns}, {Raaf}, {Stone}, {Wan}, {Wester}, {Sussman}, {Berkman},
  {Bian}, {Griskevich}, {Kropp}, {Locke}, {Mine}, {Smy}, {Sobel}, {Takhistov},
  {Weatherly}, {Ganezer}, {Hill}, {Kim}, {Lim}, {Park}, {Bodur}, {Scholberg},
  {Walter}, {Coffani}, {Drapier}, {Gonin}, {Imber}, {Mueller}, {Paganini},
  {Ishizuka}, {Nakamura}, {Jang}, {Choi}, {Learned}, {Matsuno}, {Litchfield},
  {Sztuc}, {Uchida}, {Wascko}, {Berardi}, {Calabria}, {Catanesi}, {Intonti},
  {Radicioni}, {De Rosa}, {Collazuol}, {Iacob}, {Ludovici}, {Nishimura}, {Cao},
  {Friend}, {Hasegawa}, {Ishida}, {Kobayashi}, {Nakadaira}, {Nakamura},
  {Oyama}, {Sakashita}, {Sekiguchi}, {Tsukamoto}, {Abe}, {Hasegawa}, {Isobe},
  {Miyabe}, {Nakano}, {Shiozawa}, {Sugimoto}, {Suzuki}, {Takeuchi}, {Ali},
  {Ashida}, {Hayashino}, {Hirota}, {Jiang}, {Kikawa}, {Mori}, {Nakamura},
  {Nakaya}, {Wendell}, {Anthony}, {McCauley}, {Pritchard}, {Tsui}, {Fukuda},
  {Itow}, {Murrase}, {Niwa}, {Taani}, {Tsukada}, {Mijakowski}, {Frankiewicz},
  {Jung}, {Li}, {Palomino}, {Santucci}, {Vilela}, {Wilking}, {Yanagisawa},
  {Fukuda}, {Harada}, {Hagiwara}, {Horai}, {Ishino}, {Ito}, {Koshio}, {Sakuda},
  {Takahira}, {Xu}, {Kuno}, {Cook}, {Wark}, {Di Lodovico}, {Molina Sedgwick},
  {Richards}, {Zsoldos}, {Kim}, {Tacik}, {Thiesse}, {Thompson}, {Okazawa},
  {Choi}, {Nishijima}, {Koshiba}, {Yokoyama}, {Goldsack}, {Martens}, {Murdoch},
  {Quilain}, {Suzuki}, {Vagins}, {Kuze}, {Okajima}, {Tanaka}, {Yoshida},
  {Ishitsuka}, {Matsumoto}, {Ohta}, {Martin}, {Nantais}, {Tanaka}, {Towstego},
  {Hartz}, {Konaka}, {de Perio}, {Chen}, {Jamieson}, {Walker}, {Minamino},
  {Okamoto}, {Pintaudi}, \& {Super-Kamiokande Collaboration}}]{simpson_2019_aa}
{Simpson}, C., {Abe}, K., {Bronner}, C., {et~al.} 2019, \apj, 885, 133,
  \dodoi{10.3847/1538-4357/ab4883}

\bibitem[{{Spruit}(2002)}]{spruit_2002_aa}
{Spruit}, H.~C. 2002, \aap, 381, 923

\bibitem[{{Stanton} \& {Murillo}(2016)}]{stanton_2016_aa}
{Stanton}, L.~G., \& {Murillo}, M.~S. 2016, \pre, 93, 043203,
  \dodoi{10.1103/PhysRevE.93.043203}

\bibitem[{{Suijs} {et~al.}(2008){Suijs}, {Langer}, {Poelarends}, {Yoon},
  {Heger}, \& {Herwig}}]{suijs_2008_aa}
{Suijs}, M.~P.~L., {Langer}, N., {Poelarends}, A.-J., {et~al.} 2008, \aap, 481,
  L87.
\newblock \doarXiv{0802.3286}

\bibitem[{{Thoul} {et~al.}(1994){Thoul}, {Bahcall}, \& {Loeb}}]{thoul_1994_aa}
{Thoul}, A.~A., {Bahcall}, J.~N., \& {Loeb}, A. 1994, \apj, 421, 828,
  \dodoi{10.1086/173695}

\bibitem[{{Timmes} \& {Swesty}(2000)}]{timmes_2000_ab}
{Timmes}, F.~X., \& {Swesty}, F.~D. 2000, \apjs, 126, 501,
  \dodoi{10.1086/313304}

\bibitem[{{Timmes} {et~al.}(2018){Timmes}, {Townsend}, {Bauer}, {Thoul},
  {Fields}, \& {Wolf}}]{timmes_2018_aa}
{Timmes}, F.~X., {Townsend}, R. H.~D., {Bauer}, E.~B., {et~al.} 2018, \apj,
  867, L30, \dodoi{10.3847/2041-8213/aae70f}

\bibitem[{{Townsend}(2019{\natexlab{a}})}]{mesasdk_linux}
{Townsend}, R.~H.~D. 2019{\natexlab{a}}, {MESA SDK for Linux}, 20190503,
  Zenodo, \dodoi{10.5281/zenodo.2669541}

\bibitem[{{Townsend}(2019{\natexlab{b}})}]{mesasdk_macos}
---. 2019{\natexlab{b}}, MESA SDK for Mac OS, 20190503,  Zenodo,
  \dodoi{10.5281/zenodo.2669543}

\bibitem[{{Townsend} {et~al.}(2018){Townsend}, {Goldstein}, \&
  {Zweibel}}]{townsend_2018_aa}
{Townsend}, R.~H.~D., {Goldstein}, J., \& {Zweibel}, E.~G. 2018, \mnras, 475,
  879, \dodoi{10.1093/mnras/stx3142}

\bibitem[{{Townsend} \& {Teitler}(2013)}]{townsend_2013_aa}
{Townsend}, R.~H.~D., \& {Teitler}, S.~A. 2013, \mnras, 435, 3406,
  \dodoi{10.1093/mnras/stt1533}

\bibitem[{{Trampedach} {et~al.}(2014){Trampedach}, {Stein},
  {Christensen-Dalsgaard}, {Nordlund}, \& {Asplund}}]{trampedach_2014_aa}
{Trampedach}, R., {Stein}, R.~F., {Christensen-Dalsgaard}, J., {Nordlund},
  {\AA}., \& {Asplund}, M. 2014, \mnras, 445, 4366,
  \dodoi{10.1093/mnras/stu2084}

\bibitem[{{Unno} {et~al.}(1989){Unno}, {Osaki}, {Ando}, {Saio}, \&
  {Shibahashi}}]{unno_1989_aa}
{Unno}, W., {Osaki}, Y., {Ando}, H., {Saio}, H., \& {Shibahashi}, H. 1989,
  {Nonradial oscillations of stars} ({Tokyo: University of Tokyo Press})

\bibitem[{van~der Walt {et~al.}(2011)van~der Walt, Colbert, \&
  Varoquaux}]{der_walt_2011_aa}
van~der Walt, S., Colbert, S.~C., \& Varoquaux, G. 2011, Computing in Science
  Engineering, 13, 22, \dodoi{10.1109/MCSE.2011.37}

\bibitem[{{van Horn}(1971)}]{van-horn_1971_aa}
{van Horn}, H.~M. 1971, in IAU Symposium, Vol.~42, White Dwarfs, ed. W.~J.
  {Luyten} ({Dordrecht: Springer}), 97

\bibitem[{{Vila}(1966)}]{vila_1966_aa}
{Vila}, S.~C. 1966, \apj, 146, 437, \dodoi{10.1086/148908}

\bibitem[{{Vincent} {et~al.}(2020){Vincent}, {Bergeron}, \&
  {Lafreni{\`e}re}}]{vincent_2020_aa}
{Vincent}, O., {Bergeron}, P., \& {Lafreni{\`e}re}, D. 2020, \aj, 160, 252,
  \dodoi{10.3847/1538-3881/abbe20}

\bibitem[{{Weidemann}(2000)}]{weidemann_2000_aa}
{Weidemann}, V. 2000, \aap, 363, 647

\bibitem[{{Willson}(2000)}]{willson_2000_aa}
{Willson}, L.~A. 2000, \araa, 38, 573, \dodoi{10.1146/annurev.astro.38.1.573}

\bibitem[{{Winget} {et~al.}(2004){Winget}, {Sullivan}, {Metcalfe}, {Kawaler},
  \& {Montgomery}}]{winget_2004_aa}
{Winget}, D.~E., {Sullivan}, D.~J., {Metcalfe}, T.~S., {Kawaler}, S.~D., \&
  {Montgomery}, M.~H. 2004, \apjl, 602, L109, \dodoi{10.1086/382591}

\bibitem[{{Yurchenko} {et~al.}(2011){Yurchenko}, {Barber}, \&
  {Tennyson}}]{yurchenko_2011_aa}
{Yurchenko}, S.~N., {Barber}, R.~J., \& {Tennyson}, J. 2011, \mnras, 413, 1828,
  \dodoi{10.1111/j.1365-2966.2011.18261.x}

\bibitem[{{Zahn}(1992)}]{zahn_1992_aa}
{Zahn}, J.-P. 1992, \aap, 265, 115

\end{thebibliography}

\end{document}